\UseRawInputEncoding 
\documentclass[preprint,12pt]{elsarticle}

\usepackage{amssymb}
\usepackage{lineno}

\usepackage{adjustbox}
\usepackage{xcolor,colortbl}
\usepackage{graphicx}
\usepackage{multirow}
\usepackage[bookmarks=false]{hyperref}
\usepackage{balance}
\usepackage{fancybox}
\usepackage{amsmath}

\usepackage{booktabs}
\def\bf{\textbf}

\def\fig {Figure~}

\def\tbl {Table~}

\def\sec {Section~}

\def\alg {Algorithm~}

\def\it{\textit}
\def\tr{\textrm}
\def\tt{\mct}

\usepackage{tcolorbox}
\usepackage{rotating}
\usepackage{longtable}

\usepackage{booktabs}

\usepackage{listings}
\usepackage{courier}
\def\tr{\textrm}

\usepackage{tikz}
\definecolor{1c1}{RGB}{188,162,6}
\definecolor{1c2}{RGB}{137,129,80}
\definecolor{1c3}{RGB}{239,167,31}
\definecolor{1c4}{RGB}{88,194,241}
\definecolor{1c5}{RGB}{6,180,188}

\newcommand*\circled[1]{\tikz[baseline=(char.base)]{
            \node[shape=circle,draw,inner sep=1pt] (char) {#1};}}

\def\bf{\textbf}

\def\fig {Figure~}

\def\tbl {Table~}

\def\sec {Section~}

\def\alg {Algorithm~}

\def\it{\textit}
\def\tr{\textrm}
\def\tt{\mct}
\newcommand{\ib}[1]{{\textbf {\textit { #1}}}}

\newcommand{\mct}[1]{{\footnotesize {\texttt {#1}}}}

\newcommand{\api}[1]{{\sf{\small{\texttt{#1}}}}}

\usepackage{paralist}
\lstdefinestyle{inlinecode}{basicstyle={\ttfamily\scriptsize\bfseries}}
\newcommand\code{\lstinline[style=inlinecode]}
\newcommand{\urls}[1]{{\scriptsize\url{#1}}}
\newcommand{\rev}[1]{\textcolor{black}{ #1}}
\newcommand{\revTwo}[1]{\textcolor{blue}{ #1}}

\usepackage{enumitem}

\newcommand{\nd}{\vspace{1mm}\noindent}

\usepackage{scalerel}

\usepackage{tcolorbox}
\usepackage{wrapfig}
\newcommand{\emt}[1]{\emph{``#1''}}
\usepackage[linesnumbered,noend, noline]{algorithm2e}

\usepackage{algpseudocode}

\usepackage[normalem]{ulem} 
\usepackage{xcolor}

\usepackage{ifthen}
\usepackage{amssymb}
\newboolean{showcomments}

\setboolean{showcomments}{true}

\ifthenelse{\boolean{showcomments}}
{\newcommand{\nbc}[3]{
 {\colorbox{#3}{\bfseries\sffamily\scriptsize\textcolor{white}{#1}}}
 {\textcolor{#3}{\sf\small$\blacktriangleright$\textit{#2}$\blacktriangleleft$}}
 }
}
{\newcommand{\nbc}[3]{}
 }

\usepackage{varwidth}					
\usepackage{adjustbox}

\journal{Journal IST}

\begin{document}

\begin{frontmatter}


\title{Mining API Usage Scenarios from Stack Overflow}



\author{Gias Uddin, Foutse Khomh, and Chanchal K Roy}

\address{McGill University, Polytechnique Montreal, and Unversity of Saskatchewan, Canada.}

\begin{abstract}
\nd\bf{Context:} APIs play a central role in software development. The seminal research of Carroll et al.~\cite{Carroll-MinimalManual-JournalHCI1987a} on 'minimal manual' and subsequent studies by
Shull et al.~\cite{Shull-InvestigatingReadingTechniquesForOOFramework-TSE2000} showed that developers prefer task-based API documentation instead of traditional hierarchical official documentation (e.g., Javadoc). The
Q\&A format in Stack Overflow offers developers an interface to ask and answer questions related to their development tasks. 

\nd\bf{Objective:} With a view to produce API documentation, we study automated techniques to mine API usage scenarios from Stack Overflow.

\nd\bf{Method:} We propose a framework to mine API usage scenarios from Stack
Overflow. Each task consists of a code example, the task description, and the
reactions of developers towards the code example. First, we present an algorithm
to automatically link a code example in a forum post to an API mentioned in the
textual contents of the forum post. Second, we generate a natural language
description of the task by summarizing the discussions around the code example.
Third, we automatically associate developers reactions (i.e., positive and negative opinions) towards the code example
to offer information about code quality. 

\nd\bf{Results:} We evaluate the algorithms
using three benchmarks. We compared the algorithms against seven baselines. 
Our algorithms outperformed each baseline. We developed an
online tool by automatically mining API usage scenarios from Stack Overflow. A
user study of 31 software developers shows that the participants preferred the
mined usage scenarios in Opiner over API official documentation. The tool is
available online at: \rev{http://opiner.polymtl.ca/}.

\nd\bf{Conclusion:} With a view to produce API documentation, we propose a framework to automatically mine API usage scenarios from Stack Overflow, supported by three novel algorithms. We evaluated the algorithms against
a total of eight state of the art baselines. We implement and deploy the framework in our proof-of-concept online tool, Opiner.

\end{abstract}


\end{frontmatter}



\section{Introduction}\label{sec:intro}
In 1987, the seminal research of Carroll et al.~\cite{Carroll-MinimalManual-JournalHCI1987a} introduced `minimal manual' by advocating 
the redesigning of traditional documentation around tasks, i.e., describe the software components within the contexts of development tasks. They observed 
that developers are more productive while using those manuals.  
Since then this format is proven to work better than the traditional 
API documentation~\cite{Cai-FrameworkDocumentation-PhDThesis2000,Rossen-SmallTalkMinimalistInstruction-CHI1990a,Meij-AssessmentMinimalistApproachDocumentation-SIGDOC1992}.
APIs (Application Programming Interfaces)
offer interfaces to reusable software
components. In 2000, Shull et al.~\cite{Shull-InvestigatingReadingTechniquesForOOFramework-TSE2000} 
compared traditional hierarchical API documentation (e.g., Javadocs) 
against example-based documentation, each example corresponding to a development task. They observed that the participants quickly moved to 
task-based documentation to complete their development tasks. However, task-based documentation format is 
still not adopted in API official documentation (e.g., Javadocs).

Indeed, despite developers' reliance on API official documentation as a major resource for 
learning and using APIs~\cite{Ponzanellu-Prompter-EMSE2016}, the documentation can often be incomplete, incorrect, 
and not usable~\cite{Uddin-HowAPIDocumentationFails-IEEESW2015}. This observation leads to the
question of how we can improve API documentation if the only people who can accomplish this
task are unavailable to do it. One potential way is to produce API documentation by leveraging the crowd~\cite{Subramanian-LiveAPIDocumentation-ICSE2014}, 
such as mining 
API usage scenarios from online Q\&A forums where developers discuss how they can complete development tasks using APIs. 
Although these kinds of solutions do not have the benefit of authoritativeness, recent research shows 
that developers leverage the reviews about APIs to determine how and whether an API
can be selected and used, as well as whether a provided code example is good enough for the task for
which it was given~\cite{Uddin-OpinerReviewAlgo-ASE2017,Uddin-SurveyOpinion-TSE2019,Lin-PatternOpinionMining-ICSE2019}. Thus, the combination of API reviews and code examples posted in the forum
posts may constitute an acceptable expedient in cases of rapid evolution or depleted development
resources, offering ingredients to on-demand task-centric API documentation~\cite{Robillard-OndemandDeveloperDoc-ICSME2017}. 

\begin{figure}[t]
\centering
	\centering
	\vspace{-25mm}
	\hspace*{-1.0cm}%
   	\includegraphics[scale=1.1]{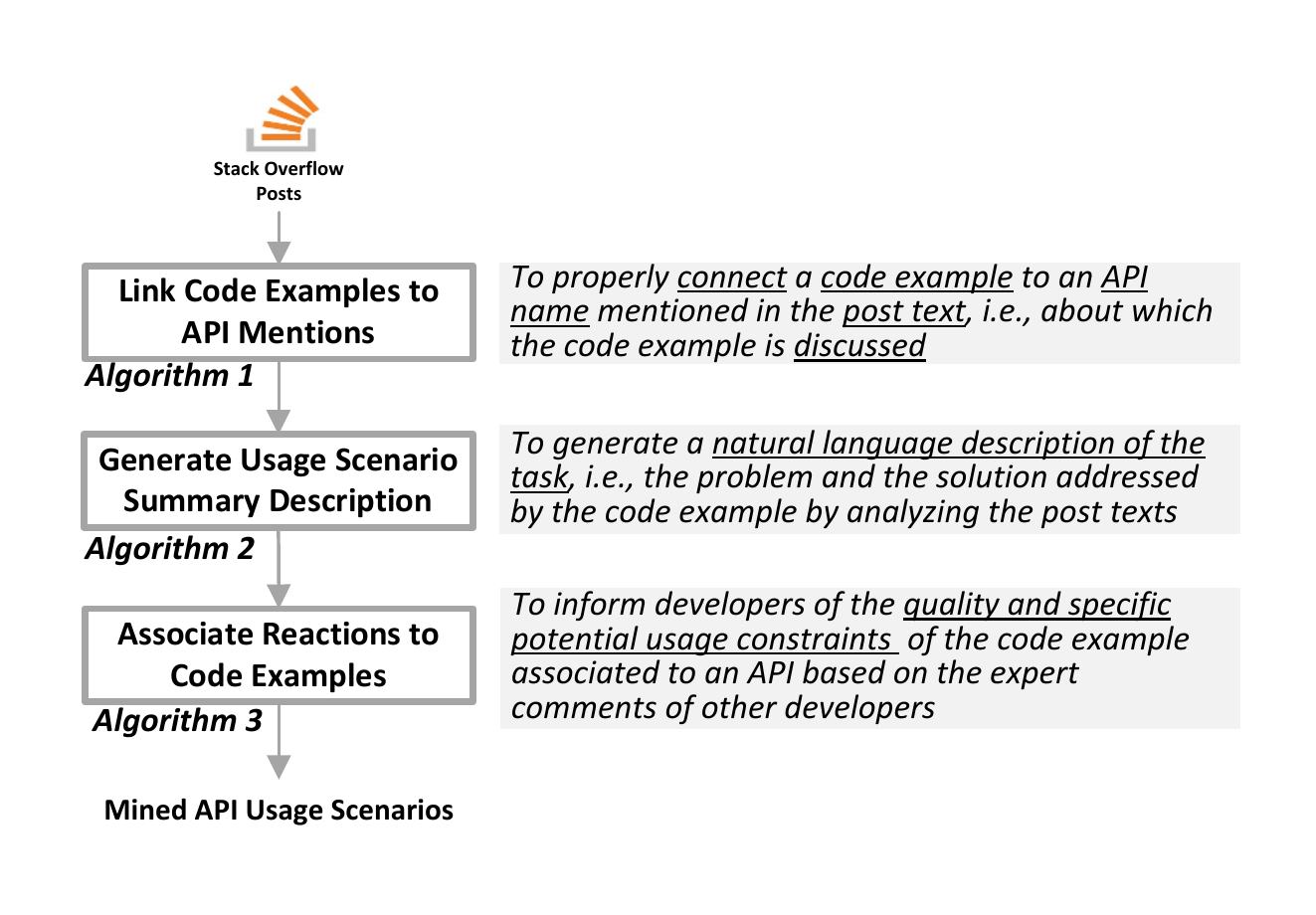}
   	\caption{Our API usage scenario mining framework from Stack Overflow with three proposed algorithms}
   	
   	 \label{fig:frameworkIntro}
\end{figure}
In this paper, with a view to assist in the automatic generation of task-based API documentation, we propose to automatically mine code examples associated to different APIs and their relevant 
task-based usage discussions from Stack Overflow. We propose an automated mining framework that can be leveraged to automatically mine API usage scenarios from 
Stack Overflow. To effectively mine API usage scenarios from Stack Overflow with high performance, we have designed and developed three algorithms within our proposed framework. 
In \fig\ref{fig:frameworkIntro}, we offer an overview of the three algorithms and show how they are used in sequence to automatically mine API usage scenarios from Stack Overflow. 

\begin{figure}
  \centering
   \hspace*{-.4cm}%
  \includegraphics[scale=1.]{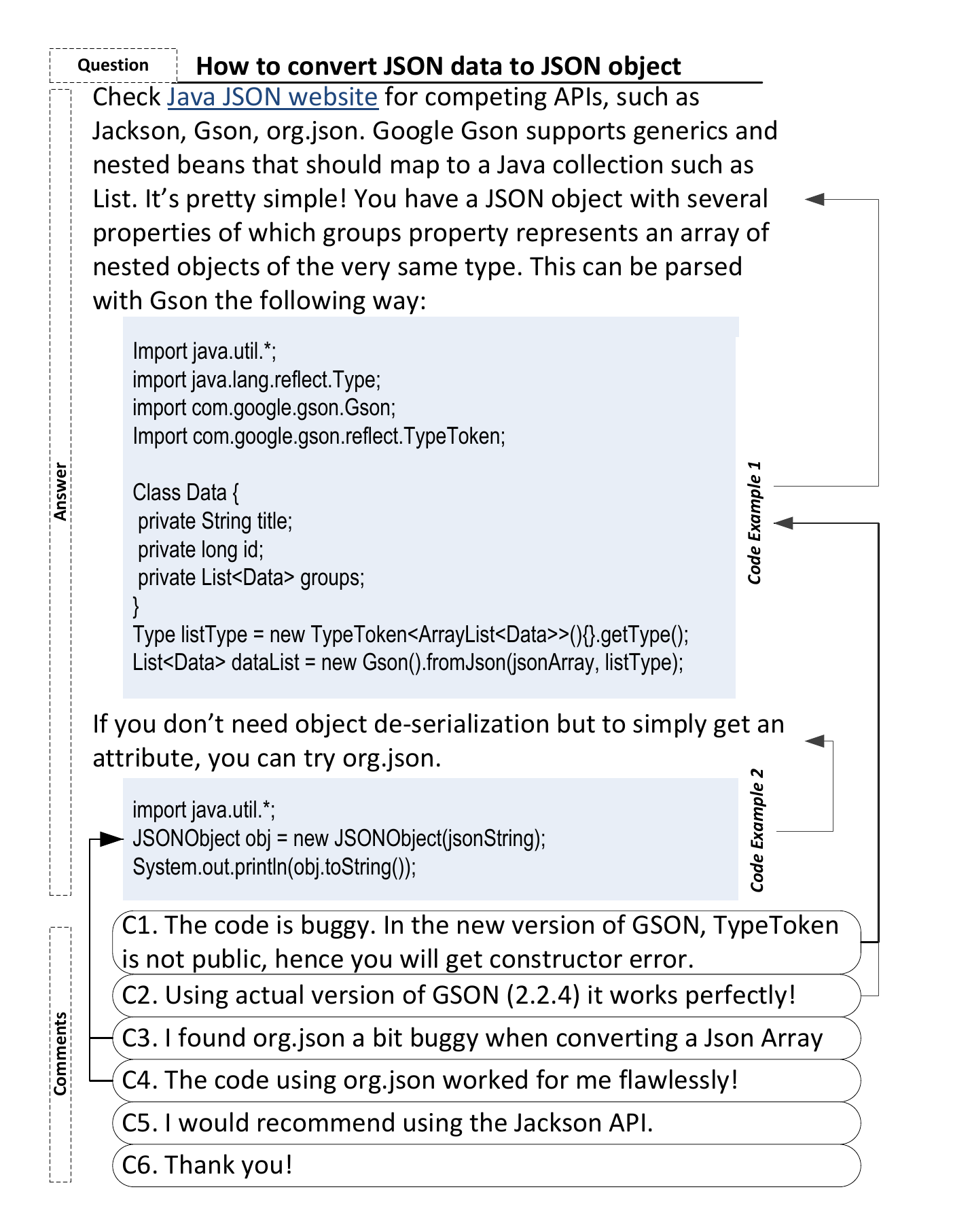}
  \caption{How API usage scenarios are discussed in Stack Overflow.
  }
 
  \label{fig:scenario-motivation}
\end{figure}

\nd\bf{$\bullet$ Algorithm 1. Associate Code Examples to API Mentions.} A code snippet is provided in a forum post to complete a development task.
Given a code snippet found in a forum post, we first need to link the snippet to an API about which the snippet is provided. 
Consider the two snippets presented in \fig\ref{fig:scenario-motivation}. Both of the snippets 
use multiple types and methods from the java.util API. In addition, the first snippet uses the java.lang API. 
However, both snippets are related to the conversion of JSON data to JSON object. As such, 
the two snippets introduce two open source Java APIs to complete the task (Google GSON in snippet 1 and org.json in snippet 2). 
The state of art traceability techniques to link code examples in 
forum posts~\cite{Subramanian-LiveAPIDocumentation-ICSE2014,Dagenais-RecoDocPaper-ICSE2012a,Phan-StatisticalLearningFQNForums-ICSE2018} 
will link the scenarios to both the utility (i.e., java.util, java.lang) and the open source APIs. 
For example, the techniques will link the first scenario to all the three APIs (java.util, java.lang,  and GSON APIs), 
even though the scenario is actually provided to discuss the usage of GSON API. This focus is easier 
to understand when we look at the textual contents that describe the usage scenario. 

\rev{Our algorithm links a code example 
to an API mentioned in the textual contents of forum post. For example, we link the first snippet in \fig\ref{fig:scenario-motivation} to 
the API GSON and the second to the API org.json. We do this by observing that both GSON and org.json are mentioned 
in the textual contents of the post, as well as the code examples consist of class and methods from the two APIs, respectively. 
We adopt the definition of an API as originally 
proposed by Martin Fowler, i.e., a ``set
of rules and specifications that a software program can follow to access and 
make use of the services and resources provided by its one or more modules''~\cite{website:wikipedia-api}. 
This definition allows us to consider a Java package as an API.   
For example, in \fig\ref{fig:scenario-motivation}, we consider the followings as APIs: \begin{inparaenum}
 \item Google GSON,
 \item Jackson,
 \item org.json,
 \item java.util, and 
 \item java.lang.
 \end{inparaenum} Each API package thus can contain a number of modules and elements (e.g., class, methods, etc.). 
 This abstraction is also consistent with the Java official documentation. For example, the \tt{java.time} packages are denoted
as the Java date APIs in the new JavaSE official tutorial~\cite{website:oracle-javadateapi}). 
As we observe in \fig\ref{fig:scenario-motivation}, this 
is also how APIs can be mentioned in online forum posts.}

%
%

\nd\bf{$\bullet$ Algorithm 2. Generate Textual Task Description.}
Given that each code snippet is provided to complete a development task, a textual description of the task as provided in 
forum posts is necessary to learn about the task as well as the underlying contexts (e.g., specific API version). 
To offer a task-based documentation for a given code snippet that is linked to an API, we made two design
decisions: \begin{inparaenum}
\item \bf{Title.} We associate each code example with the title of the question, e.g., the title
of a thread in Stack Overflow.
\item \bf{Description.} We associate relevant texts from both answer (where the code example is found) and question posts. 
For example, in \fig\ref{fig:scenario-motivation}, the first sentence (``check website \ldots'') is not important to 
learn about the tasks (i.e., JSON parsing). However, for the first snippet, all the other sentences before 
snippet 1 are necessary to learn about the 
solution (because they are all related to the API GSON that is linked to snippet 1). 
In addition, the problem description as addressed by the task can be found in the question title and post.  
\rev{Therefore, our algorithm takes as input all the texts from answer and question posts and 
outputs a summary of those textual contents based on an adaptation of the popular 
TextRank~\cite{Mihalcea-Textrank-EMNLP2004} algorithm. As explained in \sec\ref{sec:framework}, 
the TextRank algorithm is based on an adaptation of 
Google PageRank algorithm, which creates a graph of nodes and edges in a graph and ranks the nodes in the graph 
based on their association with other nodes. In our algorithm, we first heuristically find sentences relevant 
to an API in the textual contents. We then further refine their relevance by creating a graph of the sentences where each 
sentence is a node. We compute association between sentences in the graph using cosine similarity. This two-stage sentence 
selection process based on TextRank is useful to identify sentences relevant to the API task description. 
Indeed, TextRank is proven to generate high quality and relevant 
textual summary~\cite{Mihalcea-Textrank-EMNLP2004}.}    
\end{inparaenum}

\nd\bf{$\bullet$ Algorithm 3. Associate Reactions to a Code Example.} As noted before reviews about APIs can be useful to 
learn about specific nuances and usage of the provided code examples~\cite{Uddin-OpinerReviewAlgo-ASE2017,Uddin-SurveyOpinion-TSE2019}.
Consider the reactions in the comments in \fig\ref{fig:scenario-motivation}. 
Out of the six comments, two (C1, C2) are associated with the first scenario and two others (C3, C4) with the second scenario. The first comment (C1) complains 
that the provided scenario is not buggy in the newer version of the GSON API. The second comment (C2) confirms that the 
usage scenario is only valid for GSON version 2.2.4. The third comment (C3) complains that the conversion of JsonArray using org.json API is a bit buggy, but the 
next comment (C4) confirms that scenario 2 (i.e., the one related to org.json API) works flawlessly.  
Given a code example, our proposed algorithm associates relevant reactions based on heuristics, such as mentions of the linked API in a reaction (e.g., In \fig\ref{fig:scenario-motivation}, 
C1 mentions the API GSON, which is linked to code snippet 1). 

\rev{We evaluated the algorithms using three benchmarks that we created based on inputs from a total of six different human coders. 
The first benchmark consists of 730 code examples from Stack Overflow forum posts, 
each manually associated with an API mentioned in the post where the code example was found. We use the first 
benchmark to evaluate our Algorithm 1, i.e., associate code examples to API mentions. A total of three coders participated in the 
benchmark creation process. We use the second benchmark to evaluate our proposed Algorithm 2, i.e., generate textual task description addressed by a code example in Stack Overflow. 
The second benchmark consists of 216 code examples out of the 730 code examples that we used 
for the first benchmark. The 216 code examples were found in answer posts in Stack Overflow. 
The natural language summary of each of the 216 code examples was manually created based on consultations from two human coders. 
We use the third benchmark to evaluate our Algorithm 3, i.e, associate positive and negative reactions to a code example. 
The third algorithm was created by manually associated all the reactions to each of the 216 code examples that we use for the second benchmark. 
A total of three human coders participated in the benchmark creation process. The first author was the first coder 
in all the three benchmarks.}   

We observed precisions of 0.96, 0.96, and 0.89 
and recalls of 1.0, 0.98, and 0.94 with the linking of a code example to an API mention, 
the produced summaries, and the association of reactions to the code examples. We compared the algorithms against seven state of the art baselines. 
Our algorithms outperformed all the baselines. We deployed the algorithms in our online tool to mine task-based documentation from Stack Overflow. 
We evaluated the effectiveness of the tool by conducting a user study of 31 developers, each completed four coding tasks using our tool,
 API official documentation, Stack Overflow, and search engine. The developers wrote more correct code in less time and less effort using our tool. 

\section{The Mining Framework}\label{sec:framework}
\begin{figure}[t]
\centering
	\centering
	\hspace*{-.6cm}%
   	\includegraphics[scale=1.]{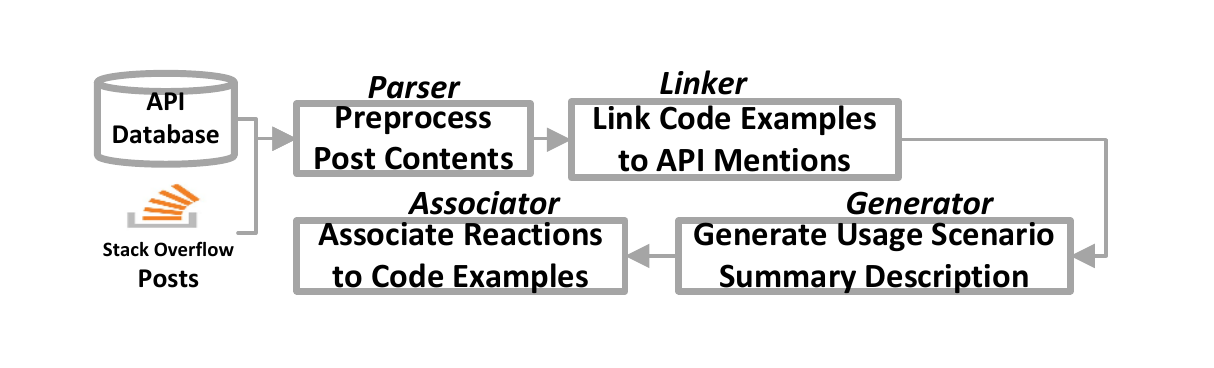}
   	\caption{The major components of our API usage scenario mining framework}
   	
   	 \label{fig:framework}
\end{figure}
We designed our framework to mine task-based API documentation by analyzing Stack Overflow, a popular forum to discuss API usage. 
The framework takes as input a forum post and outputs the usage
scenarios found in the post. For example, given as input the forum post in \fig\ref{fig:scenario-motivation}, 
the framework returns two task-based API usage scenarios:
\begin{inparaenum}[(1)]
\item The code example 1 by associating it to the API Google GSON, the two comments (C1, C2) as reactions, and a description of the code example in natural language to inform of the specific development task addressed by the code example.
\item The code example 2 by associating it to the API org.json, the two comments (C3, C4) as reactions, and a summary description.
\end{inparaenum}

\nd Our framework consists of five major components (\fig\ref{fig:framework}):

\begin{enumerate}
  \item An \bf{API database} to identify the API mentions.
  \item A suite of \bf{Parsers} to preprocess the forum post contents.
  \item A \bf{Linker} to associate a code example to an API mention.
  \item A \bf{Generator} to produce a textual task description.
  \item An \bf{Associator} to find reactions towards code examples.
\end{enumerate}


\subsection{API Database}
An API database is required to infer the association between a code 
example and an API mentioned in forum post text.
Our database consists of open source and official Java APIs.
An open-source API is identified by a name. An API
consists of one or more modules.
Each module can have one or more packages. Each package contains code elements (class, method). 
\rev{As noted in \sec\ref{sec:intro}, we consider an official Java package as an API.} 
For each API, we record the 
following meta-information: \begin{inparaenum}[(1)]
\item the name of the API,
\item the dependency of the API on other APIs,
\item the names of the modules of the API,
\item the package names under each module, 
\item the type names under each package, and 
\item the method names under each type.
\end{inparaenum} The 
last three items (package, type, and method names) can be collected from either 
the binary file of an API (e.g., a jar) or the Javadoc of the API. We obtained the first three items 
from the pom.xml files of the open-source APIs hosted in online Maven Central repository. 
Maven Central is the primary source for hosting and searching for
Java APIs with over 70 million downloads every
week~\cite{website:mavencentral-blog}.  

\subsection{Preprocessing of Forum Posts} \label{subsec:hybrid-parser}
Given as input a forum post, we preprocess its content as follows:
\begin{inparaenum}[(1)]
\item We categorize the post content into two types:
\begin{inparaenum}[(a)] \item \it{code snippets};
 \footnote{We detect code snippets as the tokens wrapped with the
 $<$code$>$ tag.}
and \item sentences in the \it{natural language text}.
\end{inparaenum}
\item Following Dagenais and Robillard~\cite{Dagenais-RecoDocPaper-ICSE2012a},
we discard the following \it{invalid} code examples based on Language-specific naming conventions:
\begin{inparaenum}[(a)]
\item Non-code snippets (e.g., XML),
\item Non-Java snippets (e.g., JavaScript).
\end{inparaenum}
\end{inparaenum} We consider the rest of the code examples as \it{valid}.

\begin{figure}[t]
  \centering
  \includegraphics[scale=.6]{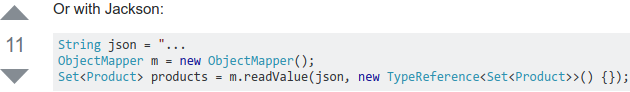}
  \vspace{-2mm}
  \caption{A popular scenario with a syntax error
  (Line 1)~\cite{website:stackoverflow-1688099}}
  \label{fig:incorrect-syntax}
\end{figure}

\nd\bf{$\bullet$ Hybrid Code Parser.} \rev{We parse each valid code snippet
using a hybrid parser combining ANTLR~\cite{Parr-ANTLR-Book2007} 
and Island Parser~\cite{Moonen-IslandParser-WCRE2001}. We observed
that code examples in the forum posts can contain syntax errors which an ANTLR parser is
not designed to parse. However, such errors can be minor and the code example
can still be useful. Consider the code example in
  \fig\ref{fig:incorrect-syntax}. An ANTLR Java parser fails at line 1 and stops there. However, the post was still considered 
  as helpful by others (upvoted 11 times). Our hybrid parser works as follows: \begin{inparaenum}
  \item We split the code example into individual lines. For this paper, we
  focused only on Java code examples. Therefore, we use
  semi-colon as the line separator indicator.
  \item We parse each line using the ANTLR parser by feeding it the
  Java grammar provided by the ANTLR package. 
If the ANTLR parser throws an exception citing parsing error,
we use our Island Parser.
  \end{inparaenum}}

\nd\bf{$\bullet$ Parsing Code Examples.} \rev{We identify API elements (types and methods) in a code example 
in three steps.}

\begin{inparaenum}[\bfseries\itshape 1.]
\item \ib{Detect API Elements:} \rev{We detect API elements 
using Java naming conventions, similar to previous 
approaches (e.g., camel
case for Class names)~\cite{Dagenais-RecoDocPaper-ICSE2012a,Rigby-CodeElementInformalDocument-ICSE2013}.
We collect types that are not declared by the user.
Consider the first code example in \fig\ref{fig:scenario-motivation}. We add \code{Type}, 
\code{Gson} and \code{TypeToken}, but not \code{Data}, 
because it was declared in the same post: \code{Class Data}.}
%
 
\rev{\item\ib{Infer Code Types From Variables:} An object instance of a
code type declared in another post can be used without any
explicit mention of the code type. For example, consider the example: \tt{Wrapper = mapper.readValue(jsonStr, Wrapper.class)}.
We associate the \tt{mapper} object to the
\tt{ObjectMapper} type, because it was defined in another post of the same thread as: \tt{ObjectMapper mapper = new ObjectMapper()}. 
}

\item\ib{Generate Fully Qualified Names (FQNs):} \rev{For each valid type
detected in the parsing, we attempt to get its fully qualified name by associating
it to an import type in the same code example. Consider the following example:}

\begin{lstlisting}
import com.restfb.json.JsonObject;
JsonObject json = new JsonObject(jsonString);
\end{lstlisting}
We associate \tt{JsonObject} to \tt{com.restfb.json.JsonObject}. We leverage both the fully and the partially qualified names in our algorithm to associate code examples to API mentions.

\end{inparaenum}

\begin{figure}[tp]
\centering
	\centering
	\hspace*{-.8cm}%
   	\includegraphics[scale=1.]{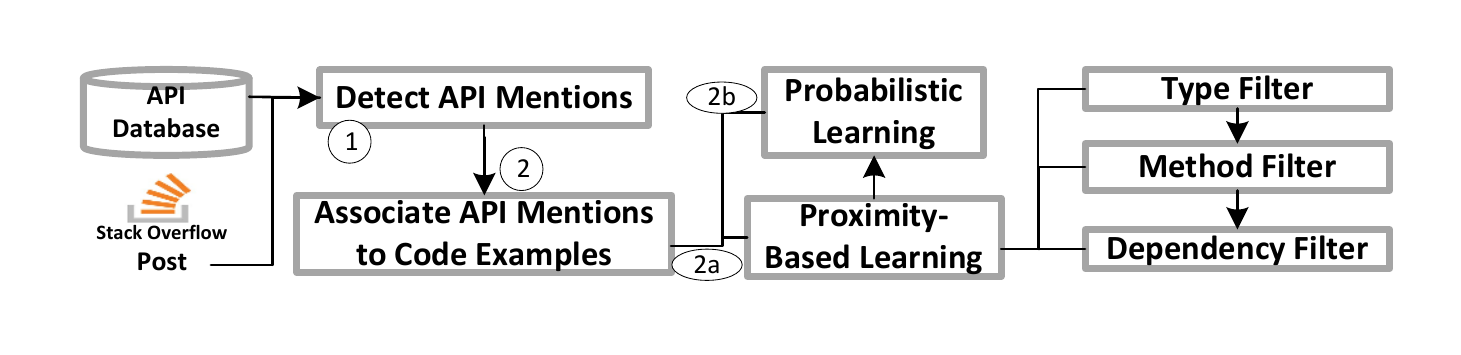}
   	\caption{The components to link a scenario to API mention}
   	
   	 \label{fig:framework-linking-algo}
\end{figure}  
 \begin{figure}[tp]
\centering
	\centering
	\hspace*{-.8cm}%
   	\includegraphics[scale=1.5]{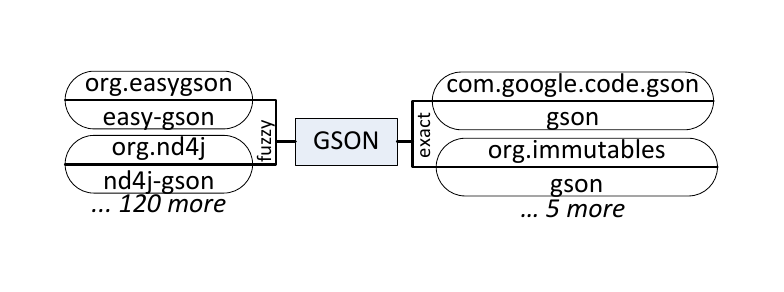}
   	\caption{Partial Mention Candidate List of GSON in \fig\ref{fig:scenario-motivation}}
   	
   	 \label{fig:mcl-gson}
\end{figure}  

\subsection{Associating Code Examples to API
Mentions}\label{subsec:association-overview} 
Given as input a code example in a forum post, we associate it to an API mentioned in the post in two steps (\fig\ref{fig:framework-linking-algo}):

\subsubsection*{Step 1. Detect API Mentions} We detect API mentions in the textual contents of forum posts following 
Uddin and Robillard~\cite{Uddin-MentionResolution-Arxiv2017}. Therefore, each API mention in our case is 
a token (or a series of tokens) if it matches at least one API or module name. Similar to~\cite{Uddin-MentionResolution-Arxiv2017}, 
we apply both exact and fuzzy matching. For example, for API mention `Gson' in \fig\ref{fig:scenario-motivation}, an exact match would be the `gson' 
module in the API `com.google.code.gson' and a fuzzy match would be the `org.easygson' API. For each such API mention, we produce a Mention Candidate List (MCL), 
by creating a list of all exact and fuzzy matches. For example, in \fig\ref{fig:mcl-gson}, we show a partial Mention Candidate List for the mention `gson'. 
Each rectangle
denotes an API candidate with its name at the top and one or more module names
at the bottom (if module names matched). 

For each code example, we create three buckets of API mentions:
\begin{inparaenum}[\bfseries (1)]
\item \it{Same Post Before $B_b$:} each mention found in the same post, but before the
code snippet. 
\item \it{Same post After $B_a$}: each mention found in the same post, but after the
code snippet.
\item \it{Same thread $B_t$}: all the mentions found in the title and in the question.
\end{inparaenum}
Each mention is accompanied by a Mention Candidate List, i.e., a list of APIs from our database.
 
\begin{algorithm}
 \SetAlgoLined
 \SetKwInOut{Input}{input}\SetKwInOut{Output}{output}
 \SetKwFunction{add}{append}
 \SetKwFunction{resolve}{getClassConf}
 \SetKwFunction{isPotentialFalse}{isPotentialFalse}
 \SetKwFunction{filter}{filter}
 \SetKwFunction{name}{name}
 \SetKwFunction{getMclTuples}{getMentionApiTuples}
 \SetKwFunction{getHits}{getHits}
 \SetKwFunction{assignUsingMaxProb}{assignUsingMaxProb}
 \SetKwFunction{Homepage}{Homepage}
  \SetKwFunction{len}{length}
    \SetKwFunction{append}{add}
    \SetKwFunction{max}{max}
 \SetKw{KwNext}{next}
 \SetKwFunction{KwMentionAPI}{MentionAPI}
 \SetKwProg{proc}{procedure}{}{}
 \SetKw{break}{break}
 \SetKw{continue}{continue}
 \SetKw{null}{null}
 \Input{
 \begin{inparaenum}[(1)]
 \item Code Example $C = (T, E)$,
 \item API Mentions in buckets $B = (B_b, B_a, B_t)$ 
 \end{inparaenum}
  
 }
 \Output{Association decision, $D = \{d_{mention}, d_{api}\}$
 }
 Proximity Filters $F = [F_{type}, F_{method}, F_{dep}]$\;
 $D = \emptyset$, $N$ = \len{$B$}, $K$ = \len{$F$}\;
 \For{$i \gets 1$ to $N$ } {
	$B_i = B[i]$, $H$ = \getMclTuples($B_i$)\;
	\For{ $k \gets 1$ to $K$} {
		Filter $F_k = F[k]$, $H$ = \getHits($F_k$, $C$, $H$, $L_i$)\;
		\lIf{$|H|$ = $1$}{
			$D = H[1]$; \break}
	}
 }
 \proc{\getMclTuples{$B$}}{
 	List$<\KwMentionAPI>$ $M$ = $\emptyset$\;
 	\ForEach{Mention $m$ $\in$ $B$} {
 		$MCL = \{a_1, a_2, \ldots a_n\}$\Comment{MCL of $m$}\;
 		\ForEach{\tr{API $a_i$ $\in$ $MCL$}} {
 			\KwMentionAPI ma = $\{m, a_i\}$; 
 			$M$.\append(ma)
 		}
 	}
 	\KwRet $M$\;
 }
 \proc{\getHits($F_k$, $C$, $H$)}{
 	$S = \emptyset$\;
 	\For{$i \gets 1$ to \len($H$) } {
 		$S[i]$ = compute score of $H[i]$ for $C$ using $F_k$\;
 	} 
 	\lIf{\max($S$) $=$ $0$} { \KwRet $H$}
 	\Else{
 		$H_{new} = \emptyset$\;
	 	\For{$i \gets 1$ to \len($H$)} {
	 		\lIf{$S[i]$  = \max($S$)}{$H_{new}$.\append($H[i]$)}
	 	}  
	 	\KwRet $H_{new}$
 	}
 } 
 \KwRet $D$
 \caption{Associate a code example to an API mention}
 \label{alg:resolution}
\end{algorithm}

\subsubsection*{Step 2. Associate Code Examples to API Mentions}
We associate a code example in a forum post to an API mention by learning how API elements in the 
code example may be connected to a candidate API in the mention candidate lists of the API mentions. We call this \it{proximity-based} learning, because we start to match with the API mentions that are more 
closer to the code example in the forum before considering the API mentions that are further away. For well-known APIs, 
we observed that developers sometimes do not mention any API name in the forum texts. In such cases, we apply \it{probabilistic learning}, by assigning the code snippet to an API that could most likely be discussed in the snippet based on the observations in other posts. 

\nd\bf{$\bullet$ Proximity-Based Learning} uses \alg\ref{alg:resolution} to associate a code example to an API mention. 
The algorithm takes as input two items: \begin{inparaenum}
\item The code example $C$, and 
\item The API mentions in the three buckets: before the code example in the post $B_b$, after the code example in the post $B_a$, 
and in the question post of the same thread $B_t$. 
\end{inparaenum}  The output from the algorithm is an association decision as a tuple ($d_{mention}$, $d_{api}$), where $d_{mention}$ is the API mention as found in the forum text (e.g., GSON for the first code example in \fig\ref{fig:scenario-motivation}) and $d_{api}$ is the name of 
the API in the mention candidate list of the API mention that is used in the code example (e.g.,  \api{com.google.code.gson} for the 
first code example in \fig\ref{fig:scenario-motivation}). 

The algorithm uses three filters (L1, discussed below). Each filter takes as input a list of tuples in the form (mention, candidate API). The output 
from the filter is a set of tuples, where each tuple in the set is ranked the highest based on the filter. The higher the ranking of a tuple, 
the more likely it is associated to the code example based on the filter. For each mention bucket (starting with $B_b$, then $B_a$, followed by $B_t$), we first create a list of tuples $H$ 
using \tt{getMentionApiTuples} (L4, L8-14). Each tuple is a pair of 
API mention and a candidate API. We apply the three filters on this list of tuples. Each filter 
produces a list of hits (L6) using \tt{getHits} procedure (L15-24). The output from a filter is passed as an input 
to the next filter, following the principle of \it{deductive learning}~\cite{Subramanian-LiveAPIDocumentation-ICSE2014}. If the list of 
hits has only one tuple, the algorithm stops and the tuple is returned as an association decision (L7). 

\ib{F1. Type Filter.} For each code type (e.g., a class)
in the code example, we search for its occurrence in the candidate APIs from Mention Candidate List. We compute type similarity between a snippet $s_i$ and a candidate $c_i$ as follows. 

{\begin{equation}
\tr{Type Similarity} = \frac{|\tr{Types}(s_i)\bigcap
\tr{Types}(c_i)|}{|\tr{Types}(s_i)|}
\end{equation}}
$\tr{Types}(s_i)$ is the list of types for $s_i$ in bucket.
$\tr{Types}(c_i)$ is the list of the types in $\tr{Types}(s_i)$ that were found in the 
types of the API. 
We associate the snippet to the API with the maximum type similarity. In case of
more than one such API, we create a \it{hit list} by putting all those APIs in
the list. Each entry is considered as a potential hit.

\ib{F2. Method Filter.} For each of candidate APIs returned in the list of hits from type filter, 
we compute method similarity between a snippet $s_i$ and a candidate
$c_i$: 

\begin{equation}
\tr{Method Similarity} = \frac{|\tr{Methods}(s_i)\bigcap
\tr{Methods}(c_i)|}{|\tr{Methods}(s_i)|}
\end{equation}
We associate the snippet to the API with the maximum similarity. In case of
more than one such API, we create a \it{hit list} of all such APIs and pass it to the next filter. 
  
\ib{F3. Dependency Filter.} We create a \it{dependency graph} by consulting the dependencies of APIs in the hit list.
Each node corresponds to an API from the hit list. An edge is established, if one API
depends on another API. From this graph, we find the API with the maximum number of incoming
edges, i.e., the API on which most of the other APIs depend on.
If there is just one such API, we assign the snippet to the API. 
This filter is developed 
based on the observation that developers mention a popular API (e.g., one on which most other APIs depend on) 
more frequently in the forum post than its dependents.    

\begin{figure}[t]
\centering
	\vspace{-2mm}
	\hspace*{-.8cm}%
  \includegraphics[scale=1.5]{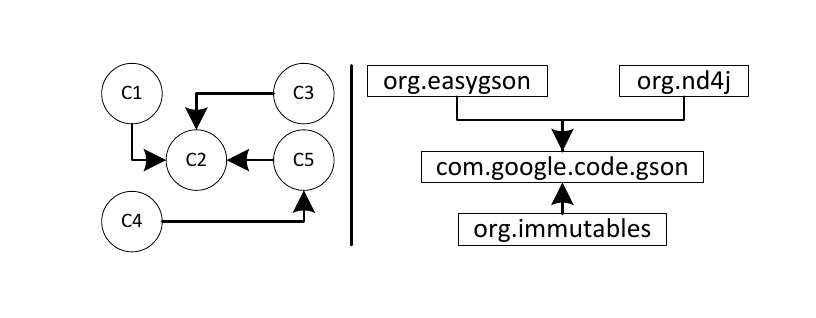}
  \vspace{-8mm}
  \caption{Dependency graph given a hit list}
  \label{fig:dep-graph}
  \vspace{-.2in}
\end{figure}

In \fig\ref{fig:dep-graph}, we show an example dependency graph (left) and a partial dependency graph for the four candidate APIs from \fig\ref{fig:mcl-gson} (right). In the left, both C2 and C5 have incoming edges, but C2 has maximum number of incoming edges. In addition, C5 depends on C2. Therefore, 
C2 is most likely the \it{core} and most popular API among the five APIs.     
The dependency filter is useful when a code example is short, with generic type and method names. In such cases, the code example 
can potentially match with many APIs. Consider a shortened version of the first code example in \fig\ref{fig:scenario-motivation}:

{\scriptsize
\begin{lstlisting}
import com.google.code.Gson; 
Data json = new Gson().fromJson(string, Data.class);
\end{lstlisting}}
Both the type (com.google.code.Gson) and methods (Gson() and fromJson(\ldots)) can be found in the two APIs in \fig\ref{fig:mcl-gson}: 
org.immutables and com.google.code.gson. However, as we see in \fig\ref{fig:dep-graph} (right), all the APIs depend on 
com.google.code.gson. Therefore, we assign the snippet to the mention Gson and the API com.google.code.gson.

\nd\bf{$\bullet$ Probabilistic Learning} is used when an API mention is not found in post texts, i.e., we cannot link a code example to an API using proximity learning. In such cases, we associate a code example to an API that was most
frequently associated in other code examples. We do this by computing the \it{coverage} of an API across those code examples linked by the proximity learning. A coverage is the total number of times the types of an API is found in those snippets. Suppose, for four code examples C1-C4, C1 and C2 are already linked to API A1, and C3 to API A2, but no API is mentioned in the post where C4 is found. In such cases, we compute the coverage of  types in C4 (say T1, T2) in the linked snippets. If T1 is present in  C1 and C2, and T2 in C3, we have coverage of 2 for API A1, and coverage of 1 for API A2. Thus, we link C4 to API A1. 
This learning is based on two observations: \begin{inparaenum}[(1)]
\item developers tend to refer
to the same API types in many different forum posts, and
\item when an API type is well-known, developers tend to refer
to it in the code examples without mentioning the API (see for
example \cite{website:stackoverflow-7141650}).
\end{inparaenum}

%


\subsection{Generating Natural Language Task Description}\label{subsec:algo-generate-task-description}
 We produce textual description for code examples that are found in the answer posts, because such a code example 
is in need to be understood for a development task~\cite{Subramanian-LiveAPIDocumentation-ICSE2014}. 
Our algorithm is based on the TextRank algorithm~\cite{Mihalcea-Textrank-EMNLP2004}. Our algorithm operates in four steps (\fig\ref{fig:summary-description-steps}):
\begin{figure}[t]
\centering
	\centering
	\hspace*{-.8cm}%
   	\includegraphics[scale=1.]{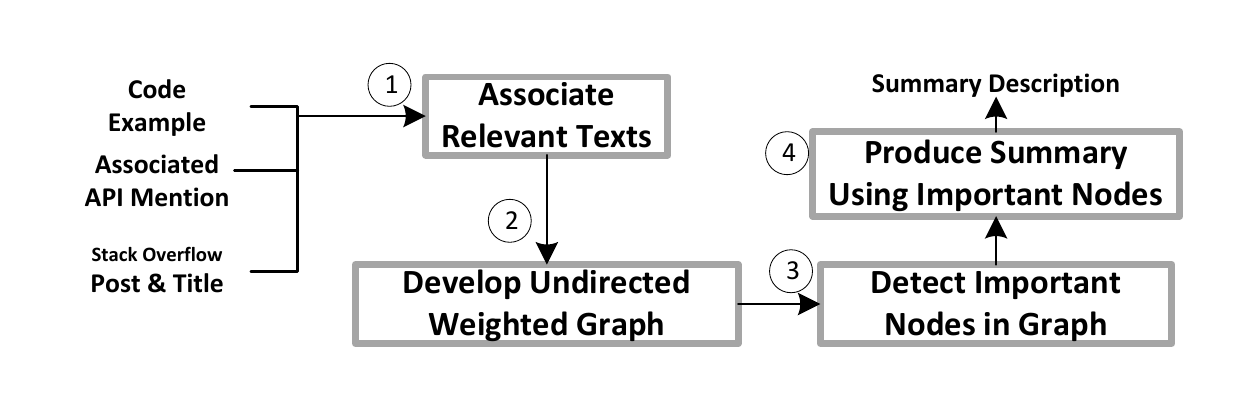}
   	\caption{Steps to produce summary description of a scenario}
   	
   	 \label{fig:summary-description-steps}
\end{figure}  

\begin{inparaenum}
\item \bf{Associate Relevant Texts.} We produce an input as a list of sentences from the forum post where the code example is found. 
Each sentence is selected by considering its proximity from the API mention linked to the code example. For example, for the first 
code example in \fig\ref{fig:scenario-motivation} linked to the API Gson, we pick all the sentences before the code example except the first one. To pick the sentences, we apply beam-search. We start with the first sentence in the forum post where API is mentioned. We then pick next possible 
sentence by looking for two types of signals: \begin{inparaenum}
\item it refers to the API (e.g., using a pronoun), and 
\item it refers to an API feature. To identify features, we use noun phrases based on shallow parsing~\cite{Klein-UnlexicalizedParsing-ACL2003a}.
\end{inparaenum} By adhering to the principle of task-oriented documentation, we organize the relevant texts into three parts: \begin{inparaenum}
\item \bf{Task Title}. The one line description of the task, as found in the title of the question. 
\item \bf{Problem.}  The relevant texts obtained from the question that describe the specific problem related to the task.
\item \bf{Solution.} The relevant texts obtained from the answer where the code example is found.
\end{inparaenum} We produce a summarized description by applying Steps 2 and 3 once for `Problem' texts and another 
for the `Solution' texts.

\item \bf{Develop Undirected Weighted Text Graph.} We remove stop words from each input sentence and then vectorize the sentence into textual units (e.g., ngram). 
We compute the distance between two sentences. A distance is defined as (1 - similarity). Similarity can be detected using standard metrics, such as 
cosine similarity. An edge is established between two sentences, if they show some similarity between them. The weight of each edge is the computed distance.  

\item \bf{Detect Important Nodes in Graph.} We traverse the text graph using the PageRank algorithm to find optimal weight for each node 
in the graph by repeatedly iterating over the following equation (until no further optimization is possible):

{
\begin{equation}
WS(V_i) = (1-d)*\sum_{V_j\in (V_i)} \frac{w_{ji}}{\sum_{v_k \in Out(V_j)} w_{jk}} WS(V_j)
\end{equation}} 
Here $d$ is the damping factor, $V$ are nodes, $WS$ are the weights. $\in (V_i)$ are the incoming edges to node $V_i$.  

\item\bf{Produce Summary Using Important Nodes.} In order to produce the summary using important nodes, we first pick the top N nodes with the most weights among all the nodes. We then rank the nodes based on their appearance in the original post (i.e., problem or solution). Each node essentially corresponds to a sentence in the post. We then combine all the ranked sentences to produce the summary. 
\end{inparaenum}

Finally, we produce a description by combining the three items in order, i.e., Title, Problem and Solution summaries.
%

\subsection{Associating Reactions to Usage Scenarios}\label{subsec:algo-associate-reaction}\label{subsec:algo-associate-reaction-scenario}
The final part of our proposed framework is to associate reactions to the usage scenarios. In order to do this, we first gather all the comments of the post where the code example is found. We then use the principles of discourage learning~\cite{liu-sentimentanalysis-handbookchapter-2010} to associate the reactions in the comments (i.e., negative and positive opinions) towards the code examples. 
The inputs to the algorithm are all the comments towards the post where the code example is found. Our algorithm works as follows. \begin{inparaenum}
\item We sort the comments in the time of posting. The earliest comment is placed at the top. We identify opinionated sentences in each comment. 
\item We identify the API mentions in each comment. 
\item We label an opinionated comment as relevant to an API mention if it refers to the API mention by name or by pronoun. To determine whether a pronoun 
refers to an API mention, we determine the distance between the API mention and the pronoun and whether another API was mentioned in between. If the opinionated 
comment is related to the API mention associated to the code example, we associate the comment to the code example. For example, in \fig\ref{fig:scenario-motivation}, the comment C4 is not considered as relevant to the code example 1, because the 
closest and most recent API name to the comment is the org.json API in comment C3.
\item For opinionated comments that do not directly/indirectly refer to an API mention (e.g., using pronoun), we associate those to the code example based on 
a notion called \it{implicit reference}. We consider a comment as implicitly related to the code example, if no other APIs are mentioned at least two comments above it.  
\end{inparaenum} 

\revTwo{To analyze the opinionated sentences, our algorithm can use the output of any sentiment detection tools.
The current framework uses an adaptation of the Domain Sentiment Orientation (DSO) algorithm as originally 
proposed by Hu et al.~\cite{Hu-MiningSummarizingCustomerReviews-KDD2004}. The algorithm was previously 
adopted by Google to analyze local service reviews~\cite{BlairGoldensohn-SentimentSummarizerLocalReviews-NLPIX2008}. 
The algorithm is called `OpnerDSO'. Given as input a sentence, the algorithm assigns it a polarity label (i.e., positive, negative, or neutral) in three steps:
\begin{enumerate}
  \item \bf{Detect potential sentiment words.} We identify adjectives in the sentence and match those against a list of sentiment words. Each word in 
  the list corresponds to either a positive or a negative polarity. The list consists of 2746 words (all adjectives) collected from three publicly available 
  datasets (the original publications of DSO~\cite{Hu-MiningSummarizingCustomerReviews-KDD2004}, MPQA~\cite{Wilson-MPQLSubjLexicon-HLT2005} 
  and AFINN~\cite{Nielsen-SentimentWordMicroblog-ESWC2011a}). 
  In addition, the list contains 750 software domain specific sentiment words that we collected by automatically crawling Stack Overflow based on two approaches, 
  Gradability~\cite{Hatzivassiloglou-adjectiveOrientationGradability-ACL2000} and Coherency~\cite{Liu-SentimentAnalysisOpinionMining-MorganClaypool2012}. Each matched 
  adjective with a positive polarity is given a score of +1 and each an adjective with a negative polarity is given a score of -1. Each score is called a sentiment orientation.      
  \item \bf{Handle negations.} We alternate the sign of a matched adjective in the presence of a negation word around the adjective, e.g., `not good' is given a score of -1 instead of +1.
  \item \bf{Label sentence.} We take the sum of all the sentiment orientations. If the sum is greater than 0, we label the sentence as `positive'. If the sum is less than 
  0, then we label it as `negative'. Otherwise, we label it as `neutral'.
\end{enumerate} We evaluated the performance of OpinerDSO on a benchmark of 4,522 sentences that we collected from 1338 Stack Overflow posts. 
A total of eight human coders labeled each sentence for polarity. 
The details of the benchmark creation process are described in~\cite{Uddin-OpinionValue-TSE2018}. We compared OpinerDSO against three sentiment 
detection tools developed for software engineering: Senti4SD~\cite{Calefato-Senti4SD-EMSE2017}, 
SentiCR~\cite{Ahmed-SentiCRNIER-ASE2017}, and SentistrengthSE~\cite{Islam-SentistrengthSE-MSR2017}. The overall precision and recall of OpinerDSO (F1-score Macro = 0.495) 
are comparable to Senti4SD (Macro F1-score = 0.510), SentiCR (Macro F1-score = 0.430), and SentistrengthSE (Macro F1-score = 0.454). Macro average is useful when we would like to
emphasize the performance on classes with few instances, which were the positive and negative polarity classes in our benchmark. The details of the evaluation are provided in~\cite{Uddin-OpinionValue-TSE2018}.}
%

\section{Evaluation}\label{sec:case-study}
We extensively evaluated the feasibility of our mining framework by investigating the accuracy of the three proposed algorithms. In particular, 
we answer the following three research questions:
\begin{enumerate}
    \item What is the performance of the algorithm to link code examples to APIs mentioned in forum texts? 
    \item What is the performance of generating the natural language summary for a scenario? 
    \item What is the performance of linking the reactions (the positive and negative opinions) to a scenario? 
\end{enumerate} 
Both high precision and recall are required in the mining of scenarios. 
A precision in the linking of a scenario to an API mention ensures we 
do not link a code example to a \it{wrong} API, a high recall ensures that we do not miss many usage scenarios 
relevant to an API. Similarly, a high precision and a high recall are required to associate reactions to a code example. %
For the summary description of a code example, a high precision is more important because otherwise we associate a wrong description 
to a code example. 

Given that all our three proposed algorithms are information retrieval in nature, we 
report four standard evaluation metrics (Precision $P$, Recall $R$, F1-score $F1$, and Accuracy $A$) as follows:  

{
\begin{eqnarray*}
P  = \frac{TP}{TP+FP},~
R = \frac{TP}{TP+FN},~
F1 = 2*\frac{P*R}{P+R},
A = \frac{TP+TN}{TP+FP+TN+FN}
\end{eqnarray*}
} $TP = $ Nb. of true positives, and $FN =$ Nb. false negatives.

\nd\bf{Evaluation Corpus.} We analyze the Stack Overflow threads tagged as `Java+JSON',
i.e., the threads contained discussions related to the JSON-based
software development tasks using Java APIs. We selected the Java JSON-based APIs because
 JSON-based techniques support diverse development scenarios, such as, both specialized
(e.g., serialization) as well as utility-based (e.g., lightweight
communication), etc. We used the `Java+JSON' threads from Stack Overflow dump of 2014 for the following reasons: 
\begin{enumerate}
\item It offers a rich set of competing APIs with
diverse usage discussions, as reported by other authors previously~\cite{Uddin-OpinerReviewAlgo-ASE2017}.
\item It 
allowed us to also check whether the API official documentation were updated with 
scenarios from the dataset (see \sec\ref{sec:discussions}). Intuitively, our mining framework is more useful when the framework can be used to 
update API official documentation by automatically mining the API usage scenarios, such as when the official documentation is found to be not updated with the 
API usage scenarios discussed in Stack Overflow even when sufficient time is spent between when such as scenario is discussed in Stack Overflow and when an API official documentation is 
last updated. 
\end{enumerate} In \tbl\ref{tbl:datasets-overview} we show descriptive statistics of the dataset.
There were 22,733 posts from 3,048 threads with scores greater than zero. Even though questions were introduced during or before 2014, each question 
is still active in Stack Overflow, i.e., the underlying tasks addressed by the questions are still relevant.  
There were 8,596
\it{valid} code snippets and 4,826 invalid code snippets. On average each
valid snippet contained at least 7.9 lines.
The last column ``Users'' show the total number of distinct users that posted at least one
answer/comment/question in those threads.

We evaluated our proposed three algorithms by creating three benchmarks out of our evaluation corpus. 
In our previous research 
of two surveys of 178 software developers, we found that  
developers consider the combination of code examples and reviews from other developers  
towards the code examples in online developer forums (e.g., Stack Overflow) as a form of API documentation. 
We also found that developers use such documentation to support diverse 
development tasks (e.g., bug fixing, API selection, feature usage, etc.)~\cite{Uddin-SurveyOpinion-TSE2017}. Therefore, 
it is necessary that our mining framework is capable of supporting any development scenario. This can be done by
 linking any code example to an API mention, 
and by producing a task-based documentation of an API to support any development task. Therefore, 
to create the benchmarks from the evaluation 
corpus, we pick code examples using random sampling that offers representation of the diverse development scenarios 
in online developer forums in general without 
focusing on a specific development scenario (e.g., How-to, bug-fixing)~\cite{Iyer-SummarizingCodeAttentionModel-ACL2016,Yin-AlignedCodeNLP-MSR2018}.  
\begin{table}[t]
\caption{Descriptive statistics of the dataset (Valid Snippets)}
\centering
\begin{tabular}{rr|rr|rr|r}\toprule
\textbf{Threads} & \textbf{Posts} & \textbf{Sentences} & \textbf{Words}  &
\textbf{Snippet} & \textbf{Lines}  & \textbf{Users}\\
\midrule
3048 & 22.7K & 87K & 1.08M & 8596  &  68.2K & 7.5K  \\ \midrule
\textbf{Average} & 7.5 & 28.6 & 353.3  & 2.8 & 7.9 & 3.9\\
\bottomrule
\end{tabular}
\label{tbl:datasets-overview}
\end{table}

\begin{table}[t]
  \centering
  \caption{Distribution of Code Snippets By APIs}
    \begin{tabular}{rrrr|rrrr}\toprule
    \multicolumn{4}{c}{\textbf{Overall}} & \multicolumn{4}{c}{\textbf{Top 5}} \\
    \midrule
    {\textbf{API}} & {\textbf{Snippet}} 
    & {\textbf{Avg}} & {\textbf{STD}} 
    & {\textbf{Snippet}} & {\textbf{Avg}} 
    & {\textbf{Max}} & {\textbf{Min}} \\
    \midrule
    175   & 8596  & 49.1  & 502.7 & 5196  & 1039.2 & 1951  & 88    \\
    \bottomrule
    \end{tabular}%
  \label{tbl:api-resolution-overview}%
\end{table}%
\begin{table}[t]
  \centering
  \caption{Distribution of Reactions in Scenarios with at least one reaction}
    \begin{tabular}{lrr|rr|rr}\toprule
    \textbf{Scenarios} & \multicolumn{2}{c}{\textbf{Comments}} & \multicolumn{2}{c}{\textbf{Positive}} & \multicolumn{2}{c}{\textbf{Negative}} \\
    \midrule
    \bf{w/reactions} & {\textbf{Total}} & {\textbf{Avg}} &{\textbf{Total}} & {\textbf{Avg}} & \multicolumn{1}{l}{\textbf{Total}} & \multicolumn{1}{l}{\textbf{Avg}} \\
    \midrule
    1154  & 7538  & 6.5   & 2487  & 2.2   & 1216  & 1.1   \\
    \bottomrule
    \end{tabular}%
  \label{tab:api-comment-overview}%
\end{table}%

\revTwo{The 8596 code examples are associated with 175
distinct APIs using our linking algorithm (see \tbl\ref{tbl:api-resolution-overview}). 
The majority (60\%) of the code examples were associated to five
APIs for parsing JSON-based files and texts in Java: java.util, org.json,
Google Gson, Jackson, and java.io. Some API types are  
more widely used in the code examples than others. For example, the \tt{Gson} class from Google Gson API was found in 679 code examples out of the 
1053 code examples linked to the API (i.e., 64.5\%). Similarly, the \tt{JSONObject} class from the org.json API was found in 
1324 of 1498 code examples linked to the API (i.e., 88.3\%). Most of those code examples also contained other types of the APIs. 
Therefore, if we follow the documentation approach of Baker~\cite{Subramanian-LiveAPIDocumentation-ICSE2014}, we would 
have at least 1324 code examples linked to the Javadoc of \tt{JSONObject} for the API org.json. This is based on the parsing of 
our 3048 Stack Overflow threads. Among the API usage scenarios in our study dataset, we found 1154 scenarios contained at least one reaction (i.e., positive or negative) using 
our proposed algorithm to associate reactions to an API usage scenario. 
In \tbl\ref{tab:api-comment-overview}, we show the distributions of comments and reactions in the 1154 scenarios. There are a total of 7,538 comments found in the corresponding 
posts of those scenarios, out of which 2,487 are sentences with positive polarity and 1,216 are sentences with negative polarity.}    

\subsection{RQ$_1$ Performance of Linking Code Example to API Mention}\label{sec:results-performance-linking}
\subsubsection{Approach} We assess the performance of our algorithm to link code examples to API mentions using a benchmark that
consists of randomly selected 730 code examples from our entire corpus.
375 code examples were
sampled from the 8589 valid code snippets and 355 from the
4826 code examples that were labeled as
invalid by the \emph{invalid code detection} component of our framework.
The
size of each subset (i.e., valid and invalid samples) is determined to capture a statistically
significant snapshot of our entire
dataset (95\% confidence interval). The evaluation corpus was manually
validated by three coders: The first two coders are the first two
authors of this paper. The third coder is a graduate student and is not a co-author.
The benchmark creation process
involved three steps:
\begin{inparaenum}[(1)]
  \item The three coders independently judged randomly
  selected 80 code examples out of the 730 code examples: 50 from the valid code
  examples and 30 from the invalid code examples. 
\item The agreement among the coders was calculated, which was
  near perfect (\tbl\ref{tbl:kappa}): pairwise Cohen $\kappa$ was 0.97 and the percent agreement was 99.4\%. To resolve disagreements on a given code example, we took the majority vote. 
 \item Since the agreement level was
 near perfect, we considered that any of the coders could complete the rest of the coding without introducing any subjective bias. The first coder then labeled the rest of the code examples.	
\end{inparaenum}
  \begin{table}[tbp]
\centering
\caption{Analysis of agreement among the coders to validate the association of APIs
to code examples (Using Recal3~\cite{recal3})}
\begin{tabular}{lrrrr}
\toprule
 & \multicolumn{1}{l}{\textbf{Kappa (Pairwise)}} &
 \multicolumn{1}{l}{\textbf{Fleiss}} & \multicolumn{1}{l}{\textbf{Percent}} & \multicolumn{1}{l}{\textbf{Krippen $\alpha$}} \\
 \midrule
\textbf{Overall} & 0.97 & 0.97 & 99.4\% & 0.97 \\
\textbf{Valid} & 0.93 & 0.93 & 98.7\% & 0.93 \\
\textbf{Discarded} & 1.0 & 1.0 & 100\% & 1.0 \\
\bottomrule
\end{tabular}
\label{tbl:kappa}
\vspace{-2mm}
\end{table}
The manual assessment found nine code examples as invalid. We labeled our algorithm as wrong for those, i.e., false positives. In the end, the benchmark consisted 
of 367 valid and 363 invalid code examples. 

\nd\bf{$\bullet$ Baselines.} We compare our algorithm against one baseline: \begin{inparaenum}[(B1)]
\item Baker~\cite{Subramanian-LiveAPIDocumentation-ICSE2014}, and 
\end{inparaenum}  
We describe the baseline below.

\bf{B1. Baker:} As noted in \sec\ref{sec:intro}, related 
techniques~\cite{Subramanian-LiveAPIDocumentation-ICSE2014,Phan-StatisticalLearningFQNForums-ICSE2018,Dagenais-RecoDocPaper-ICSE2012a} 
find fully qualified names of the API elements in the code examples. Therefore,
if a code example contains code elements from multiple APIs, 
the techniques link the code example to all APIs. We compare our algorithm 
against Baker, because it is the state of the art technique to leverage an API database in the linking 
process (unlike API usage patterns~\cite{Phan-StatisticalLearningFQNForums-ICSE2018}). 
\rev{Given that Baker was not specifically designed 
to address the type of problem we attempt to address in this
paper, we analyze both the originally proposed algorithm of Baker as well as an enhanced version of the algorithm to ensure fair comparison.} 

\begin{description}
\item[Baker (Original).] \rev{We apply the original version of the Baker algorithm~\cite{Subramanian-LiveAPIDocumentation-ICSE2014} 
on our benchmark dataset as follows.} 

\rev{\begin{inparaenum}
\item Code example consisting of code elements (type, method) only from one API: We attempt to link it using 
the technique proposed in Baker~\cite{Subramanian-LiveAPIDocumentation-ICSE2014}.
\item Code example consisting of code elements from more than one API: if the code example is associated 
to one of the API mentioned in the post, we leave it as `undecided' by Baker. 
\end{inparaenum}}
\item[Baker (Major API).] \rev{For the `undecided' API mentions by Baker (Original), we further attempt to link an API as follows. 
For a code example where Baker (original) could not decide to link it to an API mention, we link it to an API that was 
used the most frequently in the code example. We do this by computing the call frequency of each API in the code example.
Suppose, we model a code example as an API call matrix $A \times T$, where $A$ stands for an API and 
$T$ stands for a type (class, method)
of the API that is reused in the code example. The cell $(A_i, T_j)$ has a value 1 if type $T_j$ from API $A_i$ 
is called in the code example. We compute the reuse frequent of each API $A_i$ using the matrix by summing the 
number of distinct calls (to different types) is made in the code example. Thus $S_i = \sum_{j=1}^mT_j$. We assign 
the code example to the API $A_i$ with the maximum $S_i$ among all APIs reused.}       
\end{description}

 \begin{table}[t]
  \centering
\caption{Performance of linking code examples to API Mentions}
    \begin{tabular}{l|rrrr}\toprule
    \textbf{Proposed Algorithm} & \multicolumn{1}{l}{\textbf{Precision}} & \multicolumn{1}{l}{\textbf{Recall}} & \multicolumn{1}{l}{\textbf{F1 Score}} & \multicolumn{1}{l}{\textbf{Acc}} \\
    \midrule
    \textbf{Detect Invalid} & -  & -  & -   & 0.97 \\
    \textbf{Link Valid w/Partial info} & 0.94  & 1.0  & 0.97   & 0.94 \\
    \textbf{Link Valid w/Full info} & 0.96  & 1.0  & 0.98   & 0.96 \\
    \textbf{Overall w/Partial Info} & 0.94  & 0.97  & 0.95   & 0.95 \\
    \textbf{Overall w/Full Info} & 0.96  & 1.0  & 0.98   & 0.96 \\
    \midrule
    \multicolumn{4}{l}{\bf{Baselines (applied to valid code examples)}} \\
    \midrule
\bf{B1a. Baker (Original)} & 0.97 &  0.49  &  0.65 & 0.48 \\
\bf{B1b. Baker (Major API)} & 0.88 &  0.66  &  0.76 & 0.61 \\
    \bottomrule
    \end{tabular}
 \label{tbl:association-accuracy}
 \vspace{-2mm}
\end{table}%
 
\subsubsection{Results} We achieved a precision of 0.96 and a recall of 1.0 using our algorithm (\tbl\ref{tbl:association-accuracy}).
A recall of 1.0 was achieved due to the greedy approach of our algorithm which attempts to find an association 
for each code example. 
The baseline Baker (Original) 
shows the best precision among all (0.97), but with the lowest recall (0.49). This level of precision 
corroborates with the precision reported by Baker on Android SDKs~\cite{Subramanian-LiveAPIDocumentation-ICSE2014}. 
The low recall is due 
to the inability of Baker to link a code example to an API mention, when more than 
one API is used in the code example. For those code examples where Baker (Original) was undecided, we further 
attempted to improve Baker to find an API that is the most 
frequently used in the code example. The Baker (Major API) baseline improves the recall of Baker (Original) from 0.49 to 
0.66. However, the precision of Baker (Major API) drops to 0.88 from 0.97. The drop in precision is due to the fact the major API is not the API for which the code example is provided. This happened due to the 
extensive usage of Java official APIs (e.g., java.util) in the code example, while the mentioned API in the textual 
content referred to an open-source API (e.g., for Jackson/org.json for JSON parsing). In some cases the major API could 
not be determined due to multiple APIs having the maximum occurrence frequency as well as the presence of 
\it{ambiguous} types in the code example. An API type is \it{ambiguous} in our case if more than API can have a type 
with the same name. For example, \tt{JSONObject} is a popular class name among more than 900 APIs in Maven central only. 
Even the combination of type and method could be ambiguous. For example, the method 
\tt{getValue} is common for a given type, such as \tt{JSONObject.getValue(\ldots)}. In such cases, the usage of API mentions 
in the textual contents offered our proposed algorithm an improvement in precision and recall over Baker. 

We report the performance of our algorithm under different settings: \begin{inparaenum}
\item \bf{Detect Invalid.} We observed an accuracy of 0.97 to detect invalid code examples.
\item \bf{Link to valid with Partial Info.} We are able to link a valid code to an API mention with a precision of 0.94 using only the type-based filter from the proximity learning and probabilistic learning. 
This experimentation was conducted to demonstrate how much performance we can achieve with minimal information about the candidate APIs. 
Recall that the type-based filter only leverages API type names, unlike a combination of API type and method names (as used by API fully qualified name inference techniques~\cite{Subramanian-LiveAPIDocumentation-ICSE2014,Dagenais-RecoDocPaper-ICSE2012a,Phan-StatisticalLearningFQNForums-ICSE2018}. Out of the two learning rules in our algorithm, Proximity learning shows better precision than Probabilistic learning (2 vs 14 wrong associations). 
\item \bf{Link to valid with Full Info.} When we used all the filters under proximity learning, the precision level was increased to 0.96 to link a valid code example to an API mention. The slight 
improvement in precision confirms previous findings that API types (and not methods) are the major indicators for such linking~\cite{Subramanian-LiveAPIDocumentation-ICSE2014,Dagenais-RecoDocPaper-ICSE2012a}.
\item\bf{Overall.} We achieved an overall precision of 0.94 and a recall of 0.97 while using partial information. 
\end{inparaenum}

Almost one-third of the misclassified associations happened due to the code
example either being written in programming languages other than Java or the
code example being invalid. The following JavaScript code snippet was
erroneously considered as valid. It was then assigned to a wrong API: \texttt{var jsonData; \$.ajax({type: `POST'})...}. 
%

Five of the misclassifications occurred due to the code examples being
very short. Short code examples lack sufficient
API types to make an informed decision.  
Misclassifications also occurred due to the API mention detector not being able
to detect all the API mentions in a forum post. For example, the following code
example~\cite{website:stackoverflow-20374351} was erroneously 
assigned to the \api{com.google.code.gson} API. However, the correct association would be the
\api{com.google.gwt} API. The forum post (answer id 20374750) contained both API mentions. However, \api{com.google.gwt} was
mentioned using an acronym GWT and the API mention detector missed it.

{\scriptsize
\begin{lstlisting}
AutoBean<Ts> b = AutoBeanUtils.getAutoBean(ts)
return AutoBeanCodex.encode(b).getPayload(); 
\end{lstlisting}}
\subsection{RQ$_2$ Performance of Producing Textual Task Description}\label{sec:result-summary-desc-informativeness}
\subsubsection{Approach} The evaluation of natural language summary description 
can be conducted in two ways~\cite{Cohan-RevisitingSummarizationEvaluation-LREC2016}: \begin{inparaenum}
\item User study: participants are asked to rate the summaries
\item Benchmark: The summaries are compared against a benchmark.
\end{inparaenum} We follow benchmark-based settings, which compare produced summaries 
are compared against those in the benchmark using metrics, 
e.g., coverage of the sentences. 

\rev{In our previous benchmark (RQ$_1$), out of the 367 valid code example, 
216 code examples were found in the answer posts. The rest of the valid 
code examples (i.e., 151) were found in the answer posts. 
We assess the performance of our summarization algorithm for the 216 code examples that 
are found in the answer posts, because each 
code example is provided in an attempt to suggest a solution to a development task and our goal 
is to create task-based documentation support for APIs.}
 
We create another benchmark by manually producing summary description for the 216 
code examples using two types of information: \begin{inparaenum}
\item the description of the task that is addressed by the code example, and 
\item the description of the solution as carried out by the code example. 
\end{inparaenum} Both of these information types can be obtained from forum posts, such as problem definition from the question post 
and solution description from the answer post. We picked sentences following principles of extractive summarization~\cite{Cohan-RevisitingSummarizationEvaluation-LREC2016} and minimal manual~\cite{Carroll-MinimalManual-JournalHCI1987a}, i.e., pick only sentences that are related to the task. Consider a task description, \emph{``I cannot convert JSON string into Java object using Gson. I have previously used Jackson for this task".} If the provided code example is linked to the API Gson, we pick the first sentence as relevant to describe the problem, but not the second sentence. A total of two human coders were used to produce the benchmark. The first coder is the first author of this paper. The second coder is a graduate student and is not a co-author of this paper. The two coders followed the following steps: \begin{inparaenum}
\item create a coding guide to determine how summaries can be produced and evaluated,
\item randomly pick $n$ code examples out of the 216 code examples,
\item produce summary description of each code example by summarizing the problem text (from question post) and the solution text (from answer post).
\item Compute the agreement between the coders. Resolve disagreements by consulting with each other.
\item Iterate the above steps until the coders agreed on at least 80\% of the description in two consecutive iterations, i.e., after that any of the coders can produce the summary description of the rest of code examples without introducing potential individual bias. 
\end{inparaenum} In total, the two coders iterated three times and achieved at least 82\% agreement in two iterations (see \tbl\ref{tab:agreementSummDesc}). In  \tbl\ref{tab:agreementSummDesc}, the number besides an iteration shows the number of code examples that were analyzed by both coders in an iteration (e.g., 30 for the third iteration). On average, each summary in the benchmark contains 5.4 sentences and 
155.5 words.

\begin{table}[t]
  \centering
  \caption{Agreement between the coders for RQ2 benchmark}
    \begin{tabular}{lrrr}\toprule
          & \textbf{Iteration 1 (5)} & \textbf{Iteration 2 (15)} & \textbf{Iteration 3 (30)} \\
          \midrule
    \textbf{Problem} & 60.0\% & 77.8\% & 87.1\% \\
    \textbf{Solution} & 60.0\% & 87.5\% & 83.3\% \\ \midrule 
    \textbf{Overall} & 60.0\% & 82.5\% & 85.2\% \\
    \bottomrule
    \end{tabular}
  \label{tab:agreementSummDesc}%
\end{table}%

\nd\bf{$\bullet$ Baselines.} We compare against four off-the-shelf extractive 
summarization algorithms~\cite{Gambhir-SurveyTextSummarization-JAI2017}:
\begin{inparaenum}[ (B1)]
\item Luhn,
\item Lexrank,
\item TextRank, and
\item Latent Semantic Analysis (LSA).
\end{inparaenum} The first three algorithms 
were previously used to summarize API reviews~\cite{Uddin-OpinerReviewAlgo-ASE2017}. The LSA algorithms are commonly used in information retrieval and software engineering both 
for text summarization and query formulation~\cite{Haiduc-QueryReformulations-ICSE2013}. 
Extractive 
summarization techniques are the most widely used automatic summarization algorithms~\cite{Gambhir-SurveyTextSummarization-JAI2017}. Our proposed algorithm utilizes the TextRank algorithm. Therefore, by applying the TextRank algorithm without the adaption that we proposed, we can estimate the impact of the 
proposed changes.

 \begin{table}[t]
  \centering
\caption{Algorithms to produce summary description}

    \begin{tabular}{l|rrrr}\toprule
    \textbf{Techniques} & \multicolumn{1}{l}{\textbf{Precision}} & \multicolumn{1}{l}{\textbf{Recall}} & \multicolumn{1}{l}{\textbf{F1 Score}} & \multicolumn{1}{l}{\textbf{Acc}} \\
    \midrule
    \textbf{Proposed Algorithm} & 0.96  & 0.98   &  0.97    &  0.98 \\
    \midrule
\bf{B1. Luhn} & 0.66 & 0.82 & 0.71 & 0.77 \\
\bf{B2. Textrank} & 0.66 & 0.83 & 0.72 & 0.77 \\
\bf{B3. Lexrank} & 0.64 & 0.81 & 0.70 & 0.76 \\
\bf{B4. LSA} & 0.65 & 0.82 & 0.71 & 0.76 \\
    \bottomrule
    \end{tabular}%
 \label{tbl:description-accuracy}
\end{table}%
\subsubsection{Results} 
\tbl\ref{tbl:description-accuracy} summarizes the performance of the algorithm and the baselines 
to produce textual task description. We achieved the best precision (0.96) and recall (0.98) using 
our proposed engine that is built on top of the TextRank algorithm. 
Each summarization algorithm takes as input the following texts:
\begin{inparaenum}
\item the title of the question, and
\item all the textual contents from both the question and the answer posts.
\end{inparaenum} By merely applying the TextRank algorithm on the input 
we achieved a precision 0.66 and a recall of 0.83 (i.e., without the improvement of selecting sentences using beam search that we suggested in our algorithm). The improvement in our 
algorithm is due to the following two reasons:
\begin{inparaenum}
\item the selection of a smaller subset out of the input texts based on the contexts
of the code example and the associated API (i.e., Step 1 in our proposed algorithm), and 
\item the separate application of our algorithm on the Problem and Solution text blocks. 
This approach was necessary, because the baselines showed lower recall 
due to their selection of un-informative texts.
\end{inparaenum} 
The TextRank algorithm is the best performer among the baselines. 

\subsection{RQ$_3$ Performance of Linking Reactions to Code Examples}\label{sec:results-performance-linking-reaction}

\subsubsection{Approach} 
We assess the performance of our
algorithm using a benchmark that is produced by manually associating reactions
towards the 216 code examples that we analyzed for RQ1 and RQ2. 
\revTwo{Our focus is to evaluate the performance of the algorithm to 
\it{correctly} associate a reaction (i.e., positive and negative opinionated sentence) to a code example. 
As such, as we noted in \sec\ref{subsec:algo-associate-reaction-scenario}, our framework supports the 
adoption of any sentiment detection tool to detect the reactions. Given that the focus of this evaluation is on 
the \it{correct} association of reactions to code examples, we need to mitigate the threats in the evaluation that 
could arise due to the inaccuracies in the detection 
of reactions by a sentiment detection tool~\cite{Novielli-ChallengesSentimentDetectionProgrammerEcosystem-SSE2015}. 
We thus manually label the polarity (positive, negative, or neutral) 
of each sentence in our benchmark following standard guidelines in the literature~\cite{Calefato-Senti4SD-EMSE2017,Islam-SentistrengthSE-MSR2017}.} 

Out of the 216 code examples in our benchmark, 68 code examples from 59 answers consisted of at
least one comment (total 201 comments).
The 201 comments had a total of 493 sentences (190 positive, 55 negative, 248
neutral).
Four coders judged the association of each reaction (i.e., positive and negative
sentences) towards the code examples.
For each reaction, we label it either 1 (associated to the code example) or 0
(non-associated). The association of each reaction to code example was assessed
by at least two coders.
The first coder (C1) is the first author, the second (C2) is a graduate student,
third (C3) is an undergraduate student, and fourth (C4) is the second author of
the paper. The second and third coders are not co-authors of this paper. The
first coder coded all the reactions. The second and third coders coded 174 and
103 reactions, respectively. For each reaction, we took the majority vote (e.g., if C2 and C3 label as 1 but C1 as 0, we took 1,
i.e., associated). The fourth coder (C4) was consulted when a majority was not
possible. This happened for 22 reactions where two coders (C1 and C2/C3) were
involved and they disagreed. The labeling was accompanied by a coding guide.
\tbl\ref{tab:agreementLinkReaction} shows the agreement among the first three
coders.
\begin{table}[t]
  \centering
  \caption{Analysis of Agreement Between Coders To Validate the Association of Reactions to Code Examples (Using Recal2~\cite{recal2})}
    \begin{tabular}{lr|rrr}\toprule
          & \textbf{Total}  & \textbf{Percent} & \textbf{Kappa (pairwise)} & \boldmath{}\textbf{Krippen $\alpha$}\unboldmath{} \\ \midrule
    \textbf{C1-C2} & 174    & 83.9\% & 0.46 & 0.45 \\
    \textbf{C2-C3} & 51    & 62.7\% & 0.12 & 0.05 \\
    \textbf{C1-C3} & 103  & 84.5\% & 0.50 & 0.51 \\
    \bottomrule
    \end{tabular}%
  \label{tab:agreementLinkReaction}%
\end{table}%

\nd\bf{$\bullet$ Baselines.} \rev{We compare against two baselines:
\begin{inparaenum}[(B1)]
\item All Comments. A code example is linked to all the comments.
\item All Reactions. A code example is linked to all the positive and negative comments.
\end{inparaenum} The first baseline offers us insights on 
how well a blind association technique without sentiment detection may work. The second 
baseline thus includes only the subset of sentences from all sentences (i.e., B1) that are either positive 
or negative. However, not all the reactions may be related to a code example. Therefore, the 
second baseline (B2) 
offers us insights on whether the simple reliance on sentiment detection would suffice or whether 
we need a more 
sophisticated contextual approach like our proposed algorithm that picks a subset of the positive and 
negative reactions out of all reactions.}

 \begin{table}[t]
  \centering
\caption{Performance of associating reactions to code examples}
    \begin{tabular}{l|rrrr}\toprule
    \textbf{Technique} & \multicolumn{1}{l}{\textbf{Precision}} & \multicolumn{1}{l}{\textbf{Recall}} & \multicolumn{1}{l}{\textbf{F1 Score}} & \multicolumn{1}{l}{\textbf{Acc}} \\
    \midrule
    \textbf{Proposed Algorithm} &  0.89 & 0.94  &  0.91   & 0.89 \\
    \midrule
    \textbf{B1. All Comments} &  0.45 & 0.84   &  0.55    &  0.45 \\
    \textbf{B2. All Reactions} & 0.74  & 0.84  &  0.78   & 0.74 \\
    \bottomrule
    \end{tabular}%
 \label{tbl:reaction-accuracy}
\end{table}%
\subsubsection{Results} We observed the best precision (0.89) and recall (0.94) using our proposed algorithm to link reactions to code examples. The baseline `All Reactions' 
shows much better precision than the other baseline, but still lower than our algorithm. 
The lower precision of the `All Reaction' is due to the presence 
of reactions in the comments that are not related to the code example. Such reactions can be 
of two types:
\begin{inparaenum}
\item Developers offer their views of competing APIs in the comments section. Such views also 
generate reactions from other developers. 
However, to use the provided code example or complete the development task using the associated API, such discussions are not relevant. 
\item Developers can also offer views about frameworks that may be using the API associated to the code example. For example, some code examples associated with Jackson API were attributed to the 
spring framework, because spring bundles the Jackson API in its framework. We observed that such discussions were often irrelevant, because to use the Jackson 
API, a developer does not need to install the Spring framework. Therefore, from the usage perspective of the snippet, such reactions are irrelevant.
\end{inparaenum}

\section{Discussion}\label{sec:discussions}
We implemented our framework in an online tool, Opiner~\cite{Uddin-OpinerEval-ASE2017}. Using the framework deployed in Opiner, a developer can search an
API by its name to see all the mined usage scenarios of the API from Stack
Overflow. We previously developed Opiner to mine positive and negative opinions about APIs from Stack Overflow. Our proposed framework in this paper extends 
Opiner by also allowing developers to search for API usage scenarios, i.e., code examples  associated to an API and their relevant usage information. 

The current version shows results from our evaluation corpus. We present the
usage scenarios by grouping code examples that use the same
types (e.g., class) of the API. As noted in \sec\ref{sec:case-study}, our
    evaluation corpus uses Stack Overflow 2014 dataset. This choice was not
    random. We wanted to see, given sufficient time, whether the usage scenarios
    in our corpus were included in the API official documentation.
    We found a total
of 8596 valid code examples linked to 175 distinct APIs in our corpus. The
majority of those (60\%) were associated to five APIs: java.util, org.json,
Gson, Jackson, java.io.
Most of the mined scenarios for those APIs were absent in their official
documentation, e.g., for Gson, only 25\% types are used in the code examples of
its official documentation, but 81.8\% of the types are discussed in our mined
usage scenarios. 
 Therefore, the automatic mining of the usage scenarios using our framework can assist 
the API authors who could not include those in the API official documentation.

\begin{figure}
  \centering
  \hspace*{-.6cm}%
  \includegraphics[scale=1.]{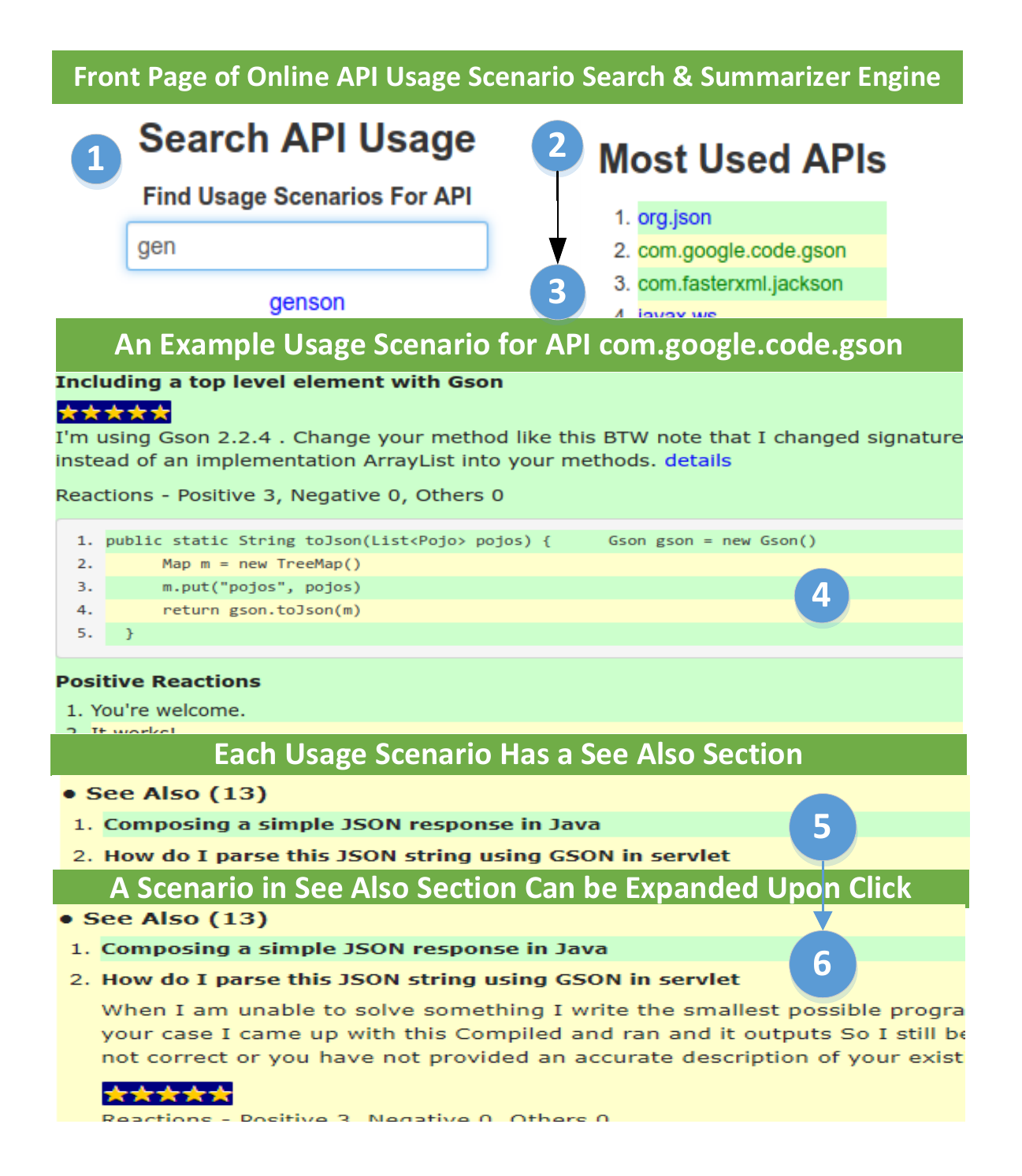}
  \hspace*{-.8cm}%
  \caption{Screenshots of online our task-based API documentation tool}
  \label{fig:opiner-intro}
  \vspace{-5mm}
\end{figure}
In \fig\ref{fig:opiner-intro}, we show screenshots of our tool.
A user can search an API by name in \circled{1} to see the mined tasks of the API
\circled{3}. An example task is shown
in \circled{4}. Other relevant tasks (i.e., that use the same classes and methods of the API) are grouped under 
`See Also' (\circled{5}). Each task under the
`See Also' can be further explored (\circled{6}). Each task is
linked to the corresponding post in Stack Overflow where the code example was
found (by clicking on the \it{details} label). The front page shows the top 10 APIs with the most mined tasks \circled{2}. 


\nd\bf{$\bullet$ Effectiveness of our Tool.}  
\rev{Although we extensively evaluated the accuracy of our algorithms, we also
measured the effectiveness of our tool with a user study. Given that
the focus of evaluation of this paper is to study the accuracy of the proposed three algorithms in our mining framework and not 
allude on the effectiveness of Opiner as a tool, we briefly 
describe the user study design and results below.}

\bf{Participants}. We recruited 31 developers. Among them, 18 were recruited through the online 
professional developers site, Freelancer.com. The other participants (13) 
were recruited from four universities, two in Canada and two in Bangladesh.
Each participant had professional software
development experience in Java. Each freelancer was remunerated with \$20. 
\revTwo{Among the 31 participants 88.2\% were actively involved in software development
(94.4\% among the freelancers and 81.3\% among the university participants).
Each participant had a background in computer science and software engineering.
The number of years of experience of the participants in software development
ranged between less than one year to more than 10 years: three (all of them being students) 
with less than one year of experience,
nine between one and two, 12 between
three and six, four between seven and 10 and the rest
(nine) had more than 10 years of experience. 
}

\bf{Tasks.} \rev{The developers each completed four coding tasks involving
four APIs (one task for each of Jackson~\cite{website:jackson},
Gson~\cite{website:gson}, Spring~\cite{website:spring} and
Xstream~\cite{website:xstream}). The four APIs were found in the list of top 10 most discussed APIs
in our evaluation corpus. The four tasks were picked randomly from our
evaluation corpus of 22.7K Stack Overflow posts. Each task was observed in
Stack Overflow posts more than once and was asked by more than one devel-
oper. Each task was related to the manipulation of JSON inputs using Java
APIs for JSON parsing. For example, the task with Jackson converts a Java object to 
JSON format, the task with Gson converts a JSON string into a Java object, the task 
with Xstream converts an XML string into a JSON string, and the task with Spring converts an 
HTTP JSON response into a Java object.}

\rev{For the user study the objects were four resources (our tool,
Stack Overflow, Official documentation, Search Engines). The participants were divided into four 
groups. Each of
first three groups (G1-3) had eight and the last group (G4) had seven participants. 
Each participant in a group was asked to complete the four coding tasks. Each
participant in a group completed the tasks using the four resources in the following order.
\begin{itemize}
  \item G1: Jackson (Stack Overflow), Gson (Javadoc), Xstream (Opiner), Spring (Everything including Search Engine)
  \item G2: Spring (Stack Overflow), Jackson (Javadoc), Gson (Opiner), Xstream (Everything including Search Engine)
  \item G3: Xstream (Stack Overflow), Spring (Javadoc), Jackson (Opiner), Gson (Everything including Search Engine)
  \item G4: Gson (Stack Overflow), Xstream (Javadoc), Spring (Opiner), Jackson (Everything including Search Engine)
\end{itemize}}

We collected the time took
to complete each task and effort spent using NASA TLX
index~\cite{Hart88} (nasatlx.com). We assessed the correctness of a solution by
computing the coverage of correct API elements. We summarize
major findings below. More details of the study the results are
provided in our online
appendix~\cite{website:opinerusagemining-online-appendix-ase2018}.

While using our tool Opiner, the participants on average coded with more correctness, 
spent the least time and effort out of all resources. 
For example, using Opiner the average time developers spent to complete a coding task was 
18.6 minutes and the average effort as reported in their TLX metrics was 45.8. 
In contrast, participants spent the highest amount of time (23.7 minutes) and effort (63.9)
per coding solution when using the official documentation. \revTwo{The difference between Opiner and official documentation 
with regards to time spent is statistically significant (p-value = 0.049) with a medium effect size. We use Mann Whitney U Test~\cite{website:scipy-mwu} to compute statistical significance, which is suitable for non-parametric testing. We use cliff's delta to compute the effect size 
and follow the effect size categorization of Romano et al.~\cite{Romano-TtestCohenD-SAIR2006}. The differences between Opiner 
and the other resources are not statistically significant for other metrics. Therefore, while the API usage scenarios in Opiner 
offer improvement over the resources, there is room for improvement.} 

After completing the tasks, 29
participants completed a survey to share their experience.
More than 80\% of the participants considered the mined usage summaries as an
improvement over both API official documentation and Stack Overflow, because our
tool offered an increase in productivity, confidence in usage and reduction in
time spent. According to one participant:\emt{It is quicker to find solution in
[tool] since the subject is well covered and useful information is collected.}
The participants considered that learning an API could be quicker while using
our tool than while using official documentation or Stack Overflow, because our
tool synthesizes the information from Stack Overflow by APIs using both
sentiment and source code analyses.

Out of the participants, 87.1\% wanted to use our tool either daily in their
development tasks, or whenever they have specific needs (e.g., learning a new
API). All the participants (100\%) rated our tool as usable  for being a single
platform to provide insights about API usage and being focused towards a
targeted audience. The developers praised the usability, search, and
analytics-driven approach in the tool. According to one participant:
\emt{In depth knowledge plus the filtered result can easily increase the
productivity of daily development tasks, \ldots with the quick glimpse of the positive and
negative feedback.} As a future improvement, the developers wished our tool to
mine usage scenarios from multiple online forums.

\section{Threats to Validity}\label{sec:threats}
\revTwo{\nd\bf{$\bullet$ External Validity} threats relate to the
generalizability of our findings and our approach. In this paper, we focus on Stack Overflow,
which is one of the largest and most popular Q\&A websites for developers. Our
findings may not generalize to other non-technical Q\&A websites that do not
focus on software development. While our evaluation corpus consists of 22.7K
posts from Stack Overflow, the results will not carry the automatic implication that the same
 results can be expected in general.}

\revTwo{\nd\bf{$\bullet$ Internal Validity} threats relate to experimenter bias and
errors while conducting the analysis. We evaluated the performance of the three proposed 
algorithms in our framework by developing three benchmarks. We mitigated the bias using manual validation (e.g., our benchmark datasets were assessed by multiple coders). 
In our user study, we assigned the study participants four tasks using for tools, including Opiner. 
Despite using a `between-subject' setting following previous research~\cite{Woh00}, the assignments were not fully counterbalanced, e.g., one out of the four groups 
had one more participant than the other groups. We compute the average of the effectiveness metrics (correctness, time, and effort spent). The absence of full counterbalance may 
still introduce some unobserved bias/error.}   

\revTwo{\nd\bf{$\bullet$ Construct Validity} threats relate to the difficulty in finding
data relevant to identify rollback edits and ambiguities. Hence, we use
revisions of the body of questions and answers from the Stack Exchange data
dump, which we think are reasonable and reliable for capturing the reasons and
ambiguities of revisions.  We also parse the web pages to create a large data
set to apply our ambiguity detection algorithms. However, we discard the
incomplete and noisy records to keep our data set clean and reliable.}

\revTwo{\nd\bf{$\bullet$ Reliability Validity} threats concern the
possibility of replicating this study. We provide the necessary data in an online appendix~\cite{website:opinerusagemining-online-appendix-ase2018}.}%
 
\section{Related Work}\label{sec:relatedWork}
Related work can broadly be divided into three areas:
\begin{inparaenum}[(1)]
\item Research in software engineering related to our three proposed algorithms,
\item Software code search tools and techniques, and 
\item crowd-sourced documentation.
\end{inparaenum}

\subsection{Works Related to the Three Proposed Algorithms} 
As we noted in \sec\ref{sec:intro}, we 
are aware of no techniques that can associate reactions towards code examples in forums (\sec\ref{subsec:algo-associate-reaction}).  

Our algorithm to generate summary description of tasks (\sec\ref{subsec:algo-generate-task-description}) 
is different from the generation of natural language description of API elements (e.g., class~\cite{Moreno-NLPJavaClasses-ICPC2013}, 
method~\cite{Sridhara-MethodSummary-ASE2010,Sridhara-DescribeActionsWithinMethods-ICSE2011}), which takes as input source code (e.g., class names, 
variable names, etc.) to produce a description. We take as input the textual discussions around code examples in forum posts. 
Our approach is different from API review 
summaries~\cite{Uddin-OpinerReviewAlgo-ASE2017}, because our summary can contain both opinionated and neutral sentences.

\rev{Our approach to generate task description from \it{an answer} 
differs from Xu et al.~\cite{BowenXu-AnswerBot-ASE2017}, who 
proposed AnswerBot to automatically summarize \it{multiple answers} relevant to a developer task. The input to AnswerBot 
is a natural language query describing a development task. Based on the query, AnswerBot first 
finds all the questions in Stack Overflow whose titles closely match the query. AnswerBot then applies 
a set of heuristics based on Maximal Marginal Relevance (MMR)~\cite{Carbonell-MMR-SIGIR1998} to find 
most novel and diverse paragraphs in the answers. The final output is the ranked order of the paragraphs 
as a summary of the answers that could be used to complete the development task. Unlike Xu et al.~\cite{BowenXu-AnswerBot-ASE2017} 
who focuses on the summarization of multiple answers for a given task, we focus on summarizing the contents of one answer. 
Unlike Xu et al.~\cite{BowenXu-AnswerBot-ASE2017} who utilize only the textual contents of answers to produce 
the summary, we utilize both the contents from the question and answer to produce the summary. 
A summary of relevant textual contents from questions provides an overview of the problem (i.e., development task).
Such a problem definition adds contextual information over the question title, which may not 
be enough to explain properly the development task. This assumption is consistent with our previous findings 
of surveys of software developers who reported the necessity of adding contextual and situationally relevant information 
into summaries produced from developer forums~\cite{Uddin-SurveyOpinion-TSE2019}.}

Our algorithm to associate a code example 
 to an API mention in a forum post (\sec\ref{subsec:association-overview}) 
  differs from the existing traceability techniques for code examples in forum posts~\cite{Subramanian-LiveAPIDocumentation-ICSE2014,Phan-StatisticalLearningFQNForums-ICSE2018,YeDeheng-ExtractAPIMentions-ICSME2016} 
  as follows:
\begin{itemize}[leftmargin=10pt]
\item \rev{As we noted in \sec\ref{sec:results-performance-linking}, the most directly comparable to our technique is Baker~\cite{Subramanian-LiveAPIDocumentation-ICSE2014}, because 
both Baker and our proposed technique rely on a predefined database of APIs. Given a code example as an input, 
our technique differs from Baker by 
considering both code examples and textual contents in the forum posts 
to learn about which API from the API database to link to the code example. Baker does not consider textual contents in the forum posts.}
\item \rev{As we noted in \sec\ref{sec:results-performance-linking}, given 
that our technique relies specifically on an API database similar to Baker~\cite{Subramanian-LiveAPIDocumentation-ICSE2014}, 
our algorithm is not directly comparable to 
StatType as proposed by Phan et al.~\cite{Phan-StatisticalLearningFQNForums-ICSE2018}. 
StatType relies on API usage patterns, i.e., how frequently a method and class name is found to be associated 
with an API in the different GitHub code repositories. 
We do not rely on the analysis of client software code to infer usage patterns of an API.}   
\item Unlike Subramanian et al.~\cite{Subramanian-LiveAPIDocumentation-ICSE2014, Dagenais-RecoDocPaper-ICSE2012a, Phan-StatisticalLearningFQNForums-ICSE2018}, 
we can operate both with \it{incomplete} and \it{complete} API database against which API mentions can be checked for traceability. 
This flexibility allowed us to use an online \it{incomplete} API database (Maven central), instead of constructing 
an offline database. All the existing traceability techniques~\cite{Subramanian-LiveAPIDocumentation-ICSE2014, Dagenais-RecoDocPaper-ICSE2012a} requires the generation of an offline \it{complete} API database to support traceability.

\item \rev{Unlike Ye et al.~\cite{YeDeheng-ExtractAPIMentions-ICSME2016}, we link a code example in a forum post to 
an API mentioned in the textual contents of the forum post. Specifically, 
Ye et al.~\cite{YeDeheng-ExtractAPIMentions-ICSME2016} focus 
on finding API methods and type names in the textual contents 
of forum posts, e.g., identify `numpy', `pandas' and `apply' in the text `While you can also use numpy, the 
documentation describes support for Pandas apply method using the following code example'. 
In contrast, our proposed algorithm links a provided code example with an API mentioned in the textual contents. 
For example, for the above textual content where Ye et al.~\cite{YeDeheng-ExtractAPIMentions-ICSME2016} link 
both `Pandas' and `numpy' APIs, our algorithm will link the provided code example to only the `Pandas' API.} 
\end{itemize} In \sec\ref{sec:results-performance-linking}, we compared our traceability algorithm with the state of the art
technique, Baker~\cite{Subramanian-LiveAPIDocumentation-ICSE2014}. The recall of Baker was 0.49, i.e., using Baker we could not
link more than 50\% code examples in our evaluation - because those contained
references to multiple API types/methods, but the textual contents referred to
only one of those APIs. Our technique could find a link for all (i.e., 100\%
recall) with more than 96\% precision. Our evaluation sample is statistically
representative of our corpus of 8589 code examples. Therefore, using Baker we
could have only found links for only 4100 of those, while our technique could
link all 8589 with a very high precision. Stack Overflow contains millions of
other code examples. Therefore, our technique significantly advances the state
of the art of code example traceability to support
task-based documentation.

\subsection{Software Code Search} 
\revTwo{Software development requires writing code to complete development tasks. Finding code examples similar to 
the task in hand can assist developers to complete the task quickly and efficiently. As such, a huge volume of research in software engineering 
has focused on the development and improvement of code search engines~\cite{Kim-Facoy-ICSE2018,Ponzanelli-PrompterRecommender-EMSE2014,
Gu-DeepCodeSearch-ICSE2018,McMillanGrechanik-Portfolio-ICSE2011,Chan-SearchingConnectedAPISubgraph-FSE2012,Haiduc-QueryReformulations-ICSE2013,
Lu-QueryExpansionViaWordnet-ICSME2015,Hill-NLBasedQueryRefinement-ICSME2014,Keivanloo-SpottingWorkingCodeExamples-ICSE2014,LvFei-CodeHow-ASE2015,Brandt-ExampleCentricProgramming-CHI2010}. 
The engines vary given the nature of input and output as well as 
the underlying searching, ranking, and visualization techniques. Based on input and output, the techniques can broadly be divided into following types: 
\begin{inparaenum}[(1)]
\item Code to code search, 
\item Code to relevant information search,
\item Natural language query to code search,
\item code snippet + natural language query to code search
\end{inparaenum}}  

\revTwo{Kim et al.~\cite{Kim-Facoy-ICSE2018} proposed FaCoy a \it{code-to-code search} engine, i.e., given as input a code snippet, 
the engine finds other code snippets that are \it{semantically} similar to the input code example. While our and FaCoY's 
goals remain the same, i.e., to help developers in their development tasks, we differ from each other with regards to both the 
outputs and the approaches. For example, given as input a code example in Stack Overflow post, we link it to an API name 
as mentioned in the textual contents of the post. In contrast, given as input a code example, FaCoY finds 
other similar code examples. 
Ponzanelli et al.~\cite{Ponzanelli-PrompterRecommender-EMSE2014} developed an Eclipse Plug-in that takes into account the source code in a given file as a context and use that to search 
Stack Overflow posts to find relevant discussions (i.e., \it{code to relevant information}). The relevant discussions are presented in a multi-faceted ranking model. In two empirical studies, Prompter's recommendations 
were found to be positive in 74\% cases. They also found that such recommendations are `volatile' in nature, since the recommendations can change at one year of distance.}

\revTwo{Natural language queries are used by leveraging text retrieval techniques to find relevant code examples (i.e., \it{natural language query to code search}). 
McMillan et al.~\cite{McMillanGrechanik-Portfolio-ICSE2011} developed Portfolio, a search engine to find relevant code functions by taking as input a natural language search query that 
offers cues of the programming task in hand. To assist in the usage of the returned functions, Portfolio also visualizes their usages.
Hill et al.~\cite{Hill-NLBasedQueryRefinement-ICSME2014} proposes an Eclipse plug-in CONQUER that takes as input a natural language query and finds relevant 
source for maintenance by incorporating multiple feedback mechanisms into the search results view, such as prevalence of the query words in the result set, etc.
Lv et al.~\cite{LvFei-CodeHow-ASE2015} proposes CodeHow, a code search technique to recognize potential APIs related to an input user query. CodeHow first attempts to understand the input query, 
and then expands the query with the potentially relevant APIs. CodeHow then performs code retrieval using the expanded query by applying a Extended Boolean Model. The model 
considers the impact of both text similarity and potential APIs during code search. In 26K C\# projects from GitHub, CodeHow achieved a precision of 0.79 based on the first 
returned relevant code snippet. In a controlled survey of Microsoft developers, CodeHow was found to be effective.
Raghothaman et al.~\cite{Raghothaman-SWIMSyntehsizingWhatIMean-ICSE2016} proposes SWIM that suggests code snippets by taking as input API-related natural language queries. 
The query does not need to contain framework specific keywords, because user query is translated into APIs of interest using click-through data from Bing search engine. 
The tool was evaluated on C\#-related queries and was found to be effective, responsive, and fast. 
Keivanloo et al.~\cite{Keivanloo-SpottingWorkingCodeExamples-ICSE2014} proposes a technique to spot working code examples among a set of code examples. The approach combines clone detection with frequent 
itemset mining~\cite{Agrawal-AssociationRuleLargeDB-ICVLD1994a} to detect popular programming solutions.  The technique 
is specifically aimed to support Internet-scale-code search engines, where usability is important. As such, the technique focuses on three criteria: \begin{inparaenum}[(1)]
\item the input and output formats should be identical to the Internet-scale search engine,
\item the query should support free-form text, and 
\item the output should be a ranked set of code examples.
\end{inparaenum}
Brandt et al.~\cite{Brandt-ExampleCentricProgramming-CHI2010} proposes to embed the results from web search engine into development environment to create a 
task-specific search engine. They propose Blueprint, a web search interface integrated into the Adobe Flex Builder developer environment to help them find locate code. 
Blueprint does this by automatically augmenting queries with code context and by presenting a code-centric view of search results. A 3-month usage logs with 2,024 users showed that 
the task-centric search interface significantly changed how and when the users searched the Web.}

\revTwo{The combination of code example/API structural knowledge and natural language queries are used to 
infer semantic meaning of the underlying programming task. Gu et al.~\cite{Gu-DeepCodeSearch-ICSE2018} proposes `Deep Code Search' by jointly embedding code snippets and natural language description into a high-dimensional 
vector space. Using the unified vector representation, code examples with purpose semantically similar to the natural language description can 
be found. The joint embedding allows semantic understanding of the purpose 
of the code and was found to be effective to retrieve relevant code snippets from large code base, such as GitHub. 
Chan et al.~\cite{Chan-SearchingConnectedAPISubgraph-FSE2012} proposes to recommend API code example by building API graph based on a simple text phrase, when the phrase may 
contain limited knowledge of the user. The API graph is buit by taking into account API invocations in client software. To optimize the traversing through the graph, they 
propose two refinement techniques. Comparison over Portfolio~\cite{McMillanGrechanik-Portfolio-ICSE2011} showed that the proposed API graph is significantly superior.} 
 
\revTwo{While natural language queries are used extensively in many code search techniques, the queries may return sub-optimal results when they are not properly formulated.
Haiduc et al.~\cite{Haiduc-QueryReformulations-ICSE2013} proposes Refoqus, a machine learning engine that takes as input a search query and recommends 
a reformulation strategy for the input query to improve its performance. The engine is trained with a sample of queries and relevant results. The tool is evaluated 
against four baseline approaches used in natural language text retrieval in five open source software systems. The tool outperformed the baselines in 84\% of the cases.
Lu et al.~\cite{Lu-QueryExpansionViaWordnet-ICSME2015} proposes to use Wordnet~\cite{Miller-WordNet-CACM1995} to expand an input natural language search query against 
a code base. The use of Wordnet allows the expansion to also include semantically relevant keywords. Evaluation against the Javascript interpreter Rhino shows that the 
synonyms derived from Wordnet to expand the queries help recommend good alternative queries.}

\revTwo{While finding relevant code example is important, an end-to-end engine can still be required to assist a developer to properly reuse the returned code example. 
Holmes et al.~\cite{Holmes-EndtoEndUserOfSourceCodeExamples-ICSME2009} outlines four case studies that involved end-to-end use of source code examples. 
The studies show that overhead and pitfalls involved in combining state-of-the-art techniques to support the use cases.}

\revTwo{While the above work return code example as an output, we return an API name as mentioned in the forum post to associate to the code example found in the same post. 
Our end goals remain the same, i.e., to assist developers to complete in their coding tasks. However, while the above work focuses on the development of a code 
search engine, we focus on the generation of an API documentation. However, it is an interesting future avenue for us to explore the possibility of extending our API traceability 
technique by combining code search techniques. For example, we could employ methods similar to FaCoY to produce a semantic representation of a code example before attempting to link it to an API.}

\subsection{Crowd-Sourced API Documentation} 
The automated mining of crowd-sourced knowledge from developer forums has generated considerable attention in recent years. 
To offer a point of reference of our analysis of related work, we review 
the research papers listed in the Stack Exchange question `Academic Papers Using Stack 
Exchange Data'~\cite{website:sopaper} and whose titles contain the keywords 
(`documentation' and/or `API')~\cite{Wang-APIsUsageObstacles-MSR2013,
Kavaler-APIsUsedinAndroidMarket-SOCINFO2013,Souza-CookbookAPI-BSSE2014,
Sunshine-APIProtocolUsability-ICPC2015,
Mastrangelo-JavaUnsafeAPIs-OOPSLA2015,
YeDeheng-ExtractAPIMentions-ICSME2016,Campos-SearchSORecommend-JSS2016,
Campos-SearchSOPostAPIBug-CASCON2016,
Azad-GenerateAPICallrules-TOSEM2017,
Ahsanuzzaman-ClassifySOPost-SANER2018,Wang2013Detecting,
Dagenais-DeveloperDocumentation-FSE2010a,
Parnin-MeasuringAPIDocumentationWeb-Web2SE2011,
Parnin2012,Jiau-FacingInequalityCrowdSourcedDocumentation-SENOTE2012,
Campbell-DeficientDocumentationDetection-MSR2013,
Treude-DocumentationInsightsSO-ICSE2016,
Delfim-RedocummentingAPIsCrowdKnowledge-JournalBrazilian2016,
LiXing-LeveragingOfficialContentSoftwareDocumentation-TSC2018,
LiSun-LearningToAnswerProgrammingQuestions-JIS2018}.    
Existing research has focused on the following areas: \begin{itemize}
\item Assessing the feasibility of forum contents for documentation and API design (e.g., usability) needs,
\item Answer question in Stack Overflow using formal documentation,
\item Recommend new documentation by complementing both official and developer forum contents, and 
\item Categorizing forum contents (e.g., detecting issues).
\end{itemize} 
Our work differs from the above work by proposing three novel algorithms that can be used to automatically generate task-based 
API documentation from Stack Overflow. As we noted in \sec\ref{sec:intro}, we follow the concept of ``minimal manual'' which promotes task-centric documentation of 
manual~\cite{Carroll-MinimalManual-JournalHCI1987a, Cai-FrameworkDocumentation-PhDThesis2000,Rossen-SmallTalkMinimalistInstruction-CHI1990a,
Meij-AssessmentMinimalistApproachDocumentation-SIGDOC1992}. 
We differ from the above work as follows: \begin{inparaenum}
\item We include comments posted in the forum as reactions to a code example in our usage scenarios. 
\item We automatically mine API usage scenarios from online forum, thereby greatly reducing the time and complexity to produce minimal manual. 
\end{inparaenum} 

Given the advance in techniques developed to automatically mine insights from crowd-sourced software 
forums, recent research on crowd-sourced API documentation has focused specifically on the analysis of quality 
in the shared knowledge. A number of high-impact recent research papers~\cite{Zhang-AreCodeExamplesInForumReliable-ICSE2018,Yang-QueryToUsableCode-MSR2016,Terragni-CSNIPPEX-ISSTA2016} 
warn against directly copying code from Stack Overflow, because such code can have potential bugs or 
misuse patterns~\cite{Zhang-AreCodeExamplesInForumReliable-ICSE2018} and that such code 
may not be directly usable (e.g., not compilable)~\cite{Yang-QueryToUsableCode-MSR2016,
Terragni-CSNIPPEX-ISSTA2016}. 
We observed both issues during the development of our proposed mining framework. We attempted to 
offer solutions to both issues within the context of our goal, i.e., producing task-based documentation. 
For example, in \sec\ref{subsec:hybrid-parser}, we discussed that shared code examples can have minor syntax problem (e.g., 
missing semicolon at the end of a source code line in Java), but they are still upvoted by Stack Overflow users, i.e., the users 
considered those code examples as useful. 
Therefore, to ensure such code examples can still be included in our task-based documentation, we developed a 
hybrid code parser that combines Island parsing with ANTLR grammar to parse code examples line by line. 
Based on the output of the parser, we thus can decide whether to include code example with syntax error or not. For example, 
if a code example has a minor error (e.g., missing semi-colon), we can decide to 
include it. We can, however, discard 
a code example that has many syntax errors (e.g., say 50\% of the source code lines have some errors).

While the issues with regards to code usability in crowd-sourced code examples~\cite{Yang-QueryToUsableCode-MSR2016,
Terragni-CSNIPPEX-ISSTA2016} could be addressed by converting those into compilable code examples, 
such approach requires 
extensive research and technological advancement due to the diversity of such issues and the 
huge number of available programming languages in modern programming environment. As a first step towards making progress in this direction, 
in our framework, we developed the algorithm to associate reactions of other developers towards a code example. The 
design and development of the algorithm was motivated by our findings from previous surveys of 178 software 
developers~\cite{Uddin-SurveyOpinion-TSE2019}. The developers reported that they consider the combination of a code example and reviews about those code examples in the forum posts as a form of
API documentation and they especially leverage the reviews to understand the potential benefits and pitfalls of reusing the code example.

\section{Conclusions}\label{sec:conclusions}
\nd\bf{$\bullet$ Summary.} APIs are central to the modern day rapid software development. However, APIs can
be hard to use due to the shortcomings in API official documentation, such as
incomplete or not usable~\cite{Robillard-APIsHardtoLearn-IEEESoftware2009a}.
This resulted in plethora of API discussions in forum posts. We present three
algorithms to automatically mine API usage scenarios from forums that can be
used to produce a task-based API documentation. We developed an online
task-based API documentation engine based on the three proposed algorithms. We evaluated the 
three algorithms using three benchmarks. Each benchmark was created by taking inputs from multiple human coders. 
We compared the algorithms against seven state-of-the-art baselines. Our proposed algorithms outperformed the baselines.

\nd\bf{$\bullet$ Future Work.} \revTwo{Our future work focuses on three major directions: 
\begin{inparaenum}[(1)]
\item The extension of the proposed framework to include all API usage scenarios from diverse developer forums, 
\item The improvement of the API usage scenario ranking in Opiner online user interface, and 
\item The utilization of the framework to produce/fix/complement traditional API documentation. 
\end{inparaenum}}


\revTwo{The ranking of API usage scenarios in Opiner website is simply based on `recency', i.e., the most recent code example is put at the top. 
This approach may not be suitable when, for example, the most recent code example is not properly described or commented. Another problem could be when 
the linking of the code example is wrong. Our future work will focus on investigating optimal ranking strategy for API usage scenarios in Opiner that can 
include both recency and additional contextual information. For example, between two most recent API usage scenarios, we can 
place the one scenario at the top that contains more description and more comments. We can also leverage the 
current research efforts to detect low quality posts in Stack Overflow during the ranking process~\cite{Ponzanelli-ClassifyQualityForumQuestion-QSIC2014,Ponzanelli-ImproveLowQualityPostDetect-ICSME2014,Ya-DetectHighQualityPosts-JIS2015,
Harper-PredictorAnswerQuality-CHI2008,Li-AnswerQualityPredictions-WWW2015,Calefato-HowToAskForTechnicalHelp-IST2018}. In addition, 
we will also focus on improving the API to code example linking accuracy.}
 
\revTwo{In our user study, the participants suggested that the usage scenarios in Opiner could be integrated into traditional API documentation. 
Given that official API documentation can be often incomplete, incorrect and obsolete~\cite{Uddin-HowAPIDocumentationFails-IEEESW2015,Robillard-APIsHardtoLearn-IEEESoftware2009a}, 
we will focus on the utilization of our proposed
framework to improve API documentation resources, such as the development of
techniques to automatically recommend fixes to common API documentation problems
(e.g., ambiguity,
incorrectness)~\cite{Uddin-HowAPIDocumentationFails-IEEESW2015,Robillard-APIsHardtoLearn-IEEESoftware2009a}, 
to associate the mined usage scenarios to specific API versions, and to produce on-demand developer
documentation~\cite{Robillard-OndemandDeveloperDoc-ICSME2017}.}

\begin{small}
\bibliographystyle{abbrv}
\bibliography{consolidated}
\end{small}

\end{document}